\documentclass[preprint, 3p,11pt,times,authoryear]{elsarticle}

\usepackage{amssymb}
\usepackage{layout}
\usepackage{bm}
\usepackage{amsmath}
\usepackage{mathtools}
\usepackage{lmodern} 
\usepackage{subcaption}

\usepackage[makeroom]{cancel}

\usepackage[figuresright]{rotating}

\journal{Journal of the Mechanics and Physics of Solids}

\newcommand{\Ba}{{\boldsymbol{\mathnormal a}}}
\newcommand{\Bb}{{\boldsymbol{\mathnormal b}}}

\newcommand{\Be}{{\boldsymbol{\mathnormal e}}}
\newcommand{\Bf}{{\boldsymbol{\mathnormal f}}}
\newcommand{\Bg}{{\boldsymbol{\mathnormal g}}}
\newcommand{\Bl}{{\boldsymbol{\mathnormal l}}}
\newcommand{\Bn}{{\boldsymbol{\mathnormal n}}}

\newcommand{\Bv}{{\boldsymbol{\mathnormal v}}}

\newcommand{\Bq}{{\boldsymbol{\mathnormal q}}}
\newcommand{\Br}{{\boldsymbol{\mathnormal r}}}

\newcommand{\Bt}{{\boldsymbol{\mathnormal t}}}
\newcommand{\BA}{{\boldsymbol{\mathnormal A}}}

\newcommand{\BL}{{\boldsymbol{\mathnormal L}}}

\newcommand{\BV}{{\boldsymbol{\mathnormal V}}}
\newcommand{\BQ}{{\boldsymbol{\mathnormal Q}}}
\newcommand{\BI}{{\boldsymbol{\mathnormal I}}}

\newcommand{\BP}{{\boldsymbol{\mathnormal P}}}
\newcommand{\BT}{{\boldsymbol{\mathnormal T}}}
\newcommand{\superscr}[1]{\ensuremath{{}^{\text #1}}}

\newcommand{\Balpha }{\ensuremath{\boldsymbol\alpha}}
\newcommand{\Bbeta  }{\ensuremath{\boldsymbol\beta}}

\newcommand{\Brho}{{\boldsymbol{\mathnormal\rho}}}

\newcommand{\Bve    }{\ensuremath{\boldsymbol\varepsilon}}

\newcommand{\vt     }{\vartheta}
\newcommand{\vp     }{\varphi}

\newcommand{\rhot   }{\ensuremath{           \rho \superscr{t}}}
\newcommand{\rhom   }{\ensuremath{           \rho \superscr{m}}}
\newcommand{\rhoep   }{\ensuremath{           \rho \superscr{e+}}}
\newcommand{\rhoem   }{\ensuremath{           \rho \superscr{e-}}}
\newcommand{\rhosp   }{\ensuremath{           \rho \superscr{s+}}}
\newcommand{\rhosm   }{\ensuremath{           \rho \superscr{s-}}}
\newcommand{\rhos   }{\ensuremath{           \rho \superscr{s}}}
\newcommand{\qt   }{\ensuremath{                q \superscr{t}}}

\newcommand{\AO     }{{\ensuremath{   {\rho} \superscr{(0)}}}}
\newcommand{\AI     }{{\ensuremath{   {\Brho} \superscr{(1)}}}}
\newcommand{\AIone}{\ensuremath{\rho_{1}^{(1)}} } 
\newcommand{\AItwo}{\ensuremath{\rho_{2}^{(1)}} } 
\newcommand{\AII     }{{\ensuremath{   {\Brho} \superscr{(2)}}}}
\newcommand{\AIIoneone}{\ensuremath{\rho_{11}^{(2)}} } 
\newcommand{\AIItwotwo}{\ensuremath{\rho_{22}^{(2)}} } 
\newcommand{\AIIonetwo}{\ensuremath{\rho_{12}^{(2)}} } 

\newcommand{\AIII     }{{\ensuremath{   {\Brho} \superscr{(3)}}}}
\newcommand{\QI     }{{\ensuremath{   {\BQ} \superscr{(1)}}}}
\newcommand{\QII     }{{\ensuremath{   {\BQ} \superscr{(2)}}}}

\newcommand{\lk}{\ensuremath{{\Bl}^{\Brho}}}
\newcommand{\lkone}{\ensuremath{{l}_1^{\Brho}}}
\newcommand{\lktwo}{\ensuremath{{l}_2^{\Brho}}}
\newcommand{\lkp}{{\ensuremath{\Bl}^{\Brho\perp}}}
\newcommand{\CDDI}{\ensuremath{{\text {CDD}}^{(1)}}}
\newcommand{\CDDII}{\ensuremath{{\text {CDD}}^{(2)}}}

\newcommand{\MI}{{\ensuremath{{M} \superscr{(1)}}}}

\renewcommand{\div}{\text{div}}
\newcommand{\Div}{\text{Div}}
\newcommand{\curl}{\text{curl}}              
\newcommand{\alphaII}{\ensuremath{\boldsymbol\alpha\superscr{I{\!}I}}}
\newcommand{\rphi}{{(\Br,\vp)}}        
        
\newcommand{\onetwo}{{\textstyle\mthin\frac{1}{2}\mthin}}          
\def\mthin{\mkern\thinmuskip}

\newcommand{\defeq}{\overset{\text{def}}{=}}

\usepackage{color}
\usepackage[normalem]{ulem} 

\newcommand{\Replace}[2]{\bgroup\noindent\textcolor{blue}{\xout{#1} #2}\egroup\ignorespacesafterend}
\newcommand{\Delete} [1]{\bgroup\noindent\textcolor{blue}{\xout{#1}}\egroup\ignorespacesafterend}
\newcommand{\Insert} [1]{\bgroup\noindent\textcolor{blue}{#1}\egroup\ignorespacesafterend}
\newcommand{\Comment}[1]{\definecolor{Mygray}{gray}{0.50}\bgroup\color{Mygray}\noindent#1\egroup\ignorespacesafterend}

\newcommand \Mehran [1] {\bgroup\noindent[\textcolor{red}{\textbf{Mehran}: #1}]\egroup\ignorespacesafterend}
\newcommand \Stefan [1] {\bgroup\noindent[\textcolor{red}{\textbf{Stefan}: #1}]\egroup\ignorespacesafterend}
\newcommand \Michael[1] {\bgroup\noindent[\textcolor{red}{\textbf{Michael}: #1}]\egroup\ignorespacesafterend}

\newcommand{\Eqref}[1]{(\ref{#1})}
\newcommand{\figref}[1]{Fig.~\ref{#1}}
\newcommand{\Figref}[1]{Fig.~\ref{#1}}
\newcommand{\secref}[1]{Section~\ref{#1}}

\begin{document}

\begin{frontmatter}


\title{Continuum Representation of Systems of Dislocation Lines: A General Method for Deriving Closed-Form Evolution Equations}


\author[]{Mehran Monavari\corref{cor1}}
\ead{mehran.monavari@fau.de}
\cortext[cor1]{Corresponding author:}

\author{Stefan Sandfeld}
\author{Michael Zaiser}
\address{Institute for Materials Simulation (WW8), Friedrich-Alexander-University Erlangen-N\"urnberg, 
Dr.-Mack-Str. 77, 90762 F\"urth, Germany}

\begin{abstract}\footnotesize
Plasticity is governed by the evolution of, in general anisotropic, systems of dislocations. We seek to faithfully represent this evolution in terms of density-like variables which average over the discrete dislocation microstructure. Starting from T. Hochrainer's continuum theory of dislocations (CDD) [Hochrainer 2015], we introduce a methodology based on the 'Maximum Information Entropy Principle' (MIEP) for deriving closed-form evolution equations for dislocation density measures of different order. These equations provide an optimum representation of the kinematic properties of systems of curved and connected dislocation lines with the information contained in a given set of density measures.  The performance of the derived equations is benchmarked against other models proposed in the literature, using discrete dislocation dynamics simulations as a reference. As a benchmark problem we study dislocations moving in a highly heterogeneous, persistent slip-band-like geometry. We demonstrate that excellent agreement with discrete simulations can be obtained in terms of a very small number of averaged dislocation fields containing information about the edge and screw components of the total and excess (geometrically necessary) dislocation densities. From these the full dislocation orientation distribution which emerges as dislocations move through a channel-wall structure can be faithfully reconstructed.  

\end{abstract}


\end{frontmatter}


\section{Introduction}

The development of physically based and predictive models for plasticity on the micro meter scale has been a challenging task ever since the connection between dislocations and the macroscopic material response has been made in the 1930s. On the one hand, continuum approaches based on naive averaging for obtaining continuous densities without taking into account the curved and line-like nature of systems of dislocations  must fail in many cases because important microstructural information is being lost; discrete approaches, on the other hand, contain full information about the microstructure. However, the number of dislocations or accumulated plastic strain can be high which can easily become a limiting factor even for today's discrete dislocation dynamics (DDD) models. Continuum models of dislocation systems which average over the discrete dislocation microstructure do not suffer from this restriction and hence might have advantages over discrete models, provided the motion of dislocations can be correctly captured in such a continuum (or dislocation-density based) framework. 

Historically, dislocation density-based plasticity models have evolved along several independent lines. The first line originates from the continuum theory of dislocations and internal stresses as developed by \citet{kroener58} and \citet{Nye1953_ActaMetall}. This theory is formulated in a geometrically rigorous manner and provides generic relationships between the dislocation microstructure, the plastic distortion and the associated internal stress fields. The fundamental object of the theory is the dislocation density tensor $\Balpha$ which is defined as the curl of the plastic distortion, $\Balpha = - \curl \Bbeta^\text{pl}$. This theory was extended by \citet{Mura1963_PhilMag} 
who formulated a kinematic equation of evolution for the dislocation density tensor, $\partial_t \Balpha = - \curl [\Bv \times \Balpha]$ where $\Bv$ is the dislocation velocity vector. In this form the theory provides a full description of dislocation microstructure evolution and of plastic deformation for situations where all dislocations are geometrically necessary dislocations (GND), i.e., where they can be envisaged as contour lines of the plastic shear strain on the respective slip systems \citep[e.g.][]{Sedlacek2003_PhilMag, Xiang2009728, Zhu2015_JMPS}. In the general case where dislocations of multiple orientations and slip systems are present, such a field theory, in order to fully capture the evolution of the dislocation microstructure, requires a spatial resolution that is well below the spacing of the individual dislocation lines  in order to make the dislocation velocity field $\Bv$ uniquely defined. (See e.g. \citet{Xia2015_MSMSE} and \citet{Zhang2015} for such implementations). A similar spatial resolution is also required for phase field approaches to dislocation microstructure evolution who directly simulate the evolution of the shear strain fields \citep{Wang2001,Rodney2003}. Such simulations, which evidently incur a high computational cost, may be envisaged as field-theoretical approaches to discrete dislocation dynamics simulation and will not be discussed in the following. Instead our focus of interest is on {\em average}, statistical descriptions of the dislocation microstructure which work in situations where dislocations of multiple orientations are present within the same volume element, such that a mapping on corresponding density-like variables provides an efficient compression of information. On such a coarse grained level, approaches which are directly based on spatial averaging of the classical dislocation density tensor $\Balpha$ can work only in exceptional circumstances: Once the fine structure of the plastic strain field is smoothed over, the dislocation lines associated with the averaged out features are no longer represented by the dislocation density tensor. Unfortunately, this background of "statistically stored" dislocations still contributes both to plastic flow and to work hardening. We are thus faced with the fundamental problem how they can be incorporated into a density-based theory. 

The second line of continuum models, which originates from the work of \citet{Johnston1959} and \citet{Kocks1976}, solved this problem by taking the radical approach of disposing with geometry altogether and considering only the "statistically stored" contribution to the dislocation density, which is characterized by a scalar density measure $\rho$ of dimension $1/$length${}^{2}$. For this density measure, phenomenological evolution equations are formulated which relate the dislocation density to the strain, $\rho = \rho(\gamma)$, and these equations were combined with other phenomenological relations which relate the scalar density of dislocations to the flow stress (e.g., the Taylor relationship $\tau=aGb\rho^{1/2}$ where $G$ is the shear modulus, $b$ the Burgers vector modulus, and $a$ a dimensionless constant) and to the plastic flow rate (e.g. the Orowan relation $\text{d}\gamma/\text{d}t=\rho bv$ where $v$ is the scalar magnitude of the dislocation velocity). This modelling approach and its derivatives were successfully used to formulate phenomenological models of work hardening and plastic flow \citep[e.g.][]{Estrin1984_AM32,Caceres2007193,Bouaziz2013}. Generally speaking, the approach works well as long as deformation is, on the scale of the description, homogeneous such that geometrically necessary dislocations -- or, equivalently, strain gradients -- need not to be taken into account%
\footnote{The limitations of the line of Gilman and Kocks can easily be seen: consider, for instance, the deformation of a polycrystal before grain boundaries yield plastically. Any approach which considers the dislocation density an increasing function of the strain will predict that dislocation densities are largest in the grain interior, where the strains have their maximum. However, it is obvious that in reality dislocation densities will be largest near the grain boundaries where dislocation motion is constrained and dislocations form pile ups which accommodate the associated strain gradients.}.

In recent years, several attempts have been made to unify both strands of continuum modeling and to arrive at models which can capture the combined evolution of "statistically stored" and "geometrically necessary" dislocation densities in a framework which can faithfully represent the underlying motion of the discrete dislocation lines. Pioneering work was done on two-dimensional (2D) systems of straight parallel dislocations by \citet{Groma1997_PhysRevB_p5807, Zaiser2001_PhysRevB, Groma2003_ActaMater} who systematically formulated evolution equations for 2D dislocation densities as statistical averages over the dynamics of the corresponding discrete dislocation systems.  Generalizing this approach to three-dimensional systems of curved dislocations has, however, proven to be quite challenging: Straight parallel dislocations can be envisaged as points in the intersecting plane, and both the mathematical definition of densities of such objects and a consistent formulation of their kinematics are almost trivial tasks. Three-dimensionally moving dislocations, on the other hand, are curved lines which move perpendicular to their line direction while remaining topologically connected, and the definition of densities of such objects and corresponding formulation of their kinematics is far from straightforward. As a consequence, models have been (and up to date still are) developed that intentionally neglect some aspects: e.g. screw-edge representations \citep{Arsenlis2004_JMPS52,Reuber2014333,Leung2015_MSMSE23} are a coarse way of approximating continuously curved dislocation loops, while other approaches rely on a geometrically consistent description of the evolution of the GND content as described by the Kr\"oner-Nye density tensor but complement this with phenomenological ad-hoc assumptions regarding the evolution of the statistically stored dislocation density and/or the contribution of such dislocations to the plastic strain rate \citep[e.g.][]{Acharya2006_JMPS,Fressengeas20113499}.  

We finally note another line of continuum dislocation theories which is based on the low-energy dislocation structure (LEDS) hypothesis proposed by \citet{Hansen1986}. The LEDS-hypothesis to a certain extent dismisses the role of the history (initial dislocation microstructure and deformation path) and indicates that among all admissible dislocation configurations the dislocation microstructure emerging under a given set of boundary constraints is the one that minimizes the energy of the crystal. Along this line \citet{Le2014} developed a continuum dislocation theory (CDT)  of straight dislocation lines.  Recently \citet{Le2016} generalized this theory to 3D curved dislocation lines (3D-CDT) using scalar density like variables that contain information about the dislocation line orientation and curvature -- an approach that has strong analogies to our higher-dimensional continuum theory discussed below. The 3D-CDT has shown promise in reproducing analytical solutions of test cases such as dislocation pile ups in simple shear deformation.  However the theory is formulated  under the assumption that ``the dislocation network is regular in the sense that nearby dislocations have nearly the same direction and orientation'' \citep{Le2016} which implies that the dislocation microstructure should locally consist of GNDs only. This assumption restricts the applicability of the theory --  in particular, it is difficult to see how it could be applied to situations where each averaging volume contains exclusively or predominantly dislocations  of zero net Burgers vector, or more generally speaking to the interplay of SSD and GND evolution which is the main focus of the present work. 

First important steps towards representing generic 3D systems of curved and connected dislocation lines through density measures date back to the work of Kosevich in the 1970s \citep{Kosevich1979_DislocationsInSolids_p33} and were later taken up by \citet{El-Azab2000_PhysRevB}. These authors introduced the idea of envisaging dislocations in a phase or state space where densities carry additional information about their line orientation. This approach was systematized and formulated in a mathematically rigorous manner by Hochrainer and co-workers who defined dislocation densities and derived their evolution in a higher-dimensional configuration space containing line orientation variables as extra dimensions \citep{Hochrainer2006_Dissertation,Hochrainer2007_PhilMag,Sandfeld2010_PhilMag90,Sandfeld2015_IJP}. This higher-dimensional continuum dislocation dynamics (hdCDD) theory is based upon statistical averaging of the kinematics of dislocations, i.e., it provides a density based representation of the manner how curved and connected lines move in a given velocity field, with results that have been demonstrated to be in quantitative agreements with those derived from the corresponding discrete dynamics \citep{Sandfeld2015_IJP}. The theory can easily handle anisotropic orientation distributions and direction dependent dislocation velocities \citep{Sandfeld2015_MRS}. The drawback is the high computational cost of implementing a dynamics on a higher-dimensional space where, in each spatial volume element, one needs to consider the evolution of the continuous orientation distribution of dislocations. 

As a consequence, attempts have been made to formulate simplified variants of the theory (all denoted as CDD) \citep{Hochrainer2009_ICNAAM,Sandfeld_JMR,Hochrainer2014JMPS,Hochrainer2015_PhilMag} which consider density-like variables that represent low-order moments of the dislocation orientation distribution only. Such CDD models were applied to a number of benchmark problems including dislocation patterning and the analysis of size effects \citep{Sandfeld_JMR,Sandfeld2011_ICNAAM,Wulfinghoff2015_IJP,Sandfeld2015_MSMSE}. \citet{Hochrainer2015_PhilMag} put these attempts on a systematic foundation by observing that the information contained in the orientation-dependent dislocation density function is exactly the same as that contained in an alignment tensor expansion of this function with respect to the orientation variables. The components of the dislocation density alignment tensors can be envisaged as density-like fields which contain more and more detailed information about the orientation distribution of dislocations. For the dislocation density alignment tensors, \citet{Hochrainer2015_PhilMag} derived an  infinite hierarchy of evolution equations where the evolution of alignment tensors of low order depends on higher-order alignment tensors. In the present paper we present a systematic method for closing this hierarchy at any given order, and we assess the performance of the resulting lowest-order CDD models. 

\emph{The goal of this paper is threefold: (i) to introduce a systematic approach for closing the evolution equations of CDD using the Maximum Information Entropy Principle (MIEP); (ii) to calculate the closure approximations for the two lowest order variants of CDD and discuss the consequences of these approximations;  (iii) to benchmark the performance of the resulting CDD models in terms of their ability to correctly represent the way how dislocations move in a strongly anisotropic dislocation microstructure.}\\

In \secref{sec:consistencies} we start by introducing the notions of kinematic and dynamic consistency as measures for the performance of averaged density-based  theories of dislocation systems. This is followed by a brief overview of the hdCDD theory in \secref{sec:theory:hdCDD}. In \secref{subsec:simplfying_HdCDD} generic steps (based on alignment tensors) for deriving simplified CDD models with 'customize-able' accuracy from hdCDD are outlined. We then turn in  \secref{subsec:DODF} to the work horse of this paper, the \emph{Maximum Information Entropy Principle} (MIEP), which will be used both as a tool for deriving CDD evolution equations, and as a means of analyzing what information is lost in simplified CDD versions as compared to hdCDD. In Sections \ref{subsec:symDODF} - \ref{subsec:CDD2} we then derive the two simplest possible CDD variants. Their performance in terms of the so-called 'kinematic consistency' (i.e. how well the evolution of curved and connected dislocations in a \emph{given} velocity field is represented) is numerically benchmarked and analyzed in \secref{sec:numerical} where we compare them to results from DDD and  hdCDD simulations, as well as to some of alternative models published in the literature. As a benchmark example we use a problem that has been extensively studied also in the experimental literature, namely the motion of dislocations in the periodic wall structure of persistent slip bands in fatigued metals.

\section{Kinematic and dynamic consistency: How to measure the quality of averaged descriptions of dislocation systems}
\label{sec:consistencies}
The first essential step in constructing an averaged continuum theory of dislocations -- even though this is rarely made explicit -- consists in the definition of statistical rules which map a discrete dislocation system (or an ensemble of such systems) onto a set of continuous density measures. Furthermore,  we need to formulate equations of evolution for the density measures which map an initial set of density measures onto a set of measures at a later time. The performance of a continuum theory may then be assessed by investigating to which extent this mapping meets the requirements of \emph{kinematic and dynamic consistency} with the underlying discrete dynamics:

The \emph{kinematic consistency} of a continuum theory of dislocation motion can be envisaged as its ability to correctly account for the fact that dislocations are curved and connected lines which evolve by moving perpendicular to their line direction. It is thus a measure of the ability of a continuum theory to still account, after averaging, for the peculiar geometrical properties of the underlying discrete objects. To assess kinematic consistency, we propose to use a benchmarking method: For benchmark problems, we construct an initial discrete configuration (or an ensemble of such configurations) and evaluate the corresponding density measures. We then impose a dislocation velocity field $v(\Br)$ where the magnitude of the dislocation velocity does not depend on the dislocation configuration - i.e., we neglect internal stresses and dislocation interactions (obtaining those would be the problem of \emph{dislocation dynamics}). We investigate in parallel the evolution of the discrete system by moving all lines with velocity $v(\Br)$ perpendicular to their direction, and the evolution of the density measures under the respective evolution equations. Finally, after some time $t$, we evaluate the density pattern resulting from the evolved discrete system and compare with the evolved continuous measure. Kinematic consistency of a continuum theory can then be assessed in terms of the degree of agreement between the evolved continuous system, and the density representation of the evolved discrete system (c.f. \citet{Sandfeld2015b_MSMSE}). 

The \emph{dynamic consistency} of a continuum theory concerns its ability to predict the actual evolution of dislocation systems under external load. Evaluation proceeds in much the same manner as for dynamic consistency, however, the dislocation velocity field is not imposed externally. Instead, a set of external boundary conditions is specified and the dislocations evolve under the influence of the associated stresses, as well as of their mutual interactions. The local dislocation velocities are computed as functionals of stress and of the variables describing the dislocation configuration. Consistency is again evaluated by comparing the evolved continuous system with the density representation of the evolved discrete system(s). 

Kinematic consistency may be considered a special case of the more general concept of dynamic consistency. The two concepts coincide in certain limiting cases where the dislocation velocity is mainly controlled by 'external' stresses while dislocation interactions are unimportant - an example being semiconductor single crystals with ultra-low dislocation densities (and thus small dislocation interaction stresses) and high Peierls barriers. In such systems we may write the dislocation velocities as  functions of the stresses arising from external boundary constraints or differential thermal expansion, without consideration of variables characterizing the dislocation microstructure. In general, however, dynamic consistency is a much more stringent and demanding requirement than kinematic consistency. While kinematic consistency merely tells us that a continuum model accounts for the geometrical nature of dislocations, dynamic consistency additionally requires that the model correctly represents the physics of dislocation interactions, i.e. the manner how internal stresses and the resulting long-range and short-range dislocation interactions govern dislocation motion. So the reader might legitimately ask why we are so obsessed with kinematics. 

There are two answers to this question. First, we insist that kinematic consistency is a necessary prerequisite for achieving dynamic consistency: Without investigating the issue of kinematic consistency we do not even know whether an averaged theory correctly captures {\em where} dislocation lines are moving (kinematic consistency), which makes it quite futile to ask the question {\em how fast} they are moving (dynamic consistency). We note that, in order to ensure kinematic consistency, it is not sufficient to ensure that the dislocation density tensor $\Balpha$, as calculated from the density measures, is divergence free (${\div}\, \Balpha = 0$). This requirement, which ensures that the geometrically necessary dislocations can be considered a set of closed contour lines of the strain field, is a necessary condition for kinematic consistency but not sufficient to ensure correct representation of the coupled kinematics of systems containing both statistically stored and geometrically necessary dislocations. Second, and possibly more important, solving the problem of dislocation kinematics is a step which can help to systematically derive theories of dislocation dynamics. In a recent paper of \citet{Hochrainer2016} it was demonstrated that, if the energy of a dislocation system is known as a functional of a set of density measures and the kinetic equations which govern the evolution of the same measures are also known, then one can use the powerful framework of linear irreversible thermodynamics to systematically construct thermodynamically consistent dynamic theories. In another recent paper by \citet{Zaiser2015_PRB} it was further demonstrated that the elastic energy functional of a discrete dislocation system can be averaged to express the elastic energy as a functional of the corresponding dislocation density alignment tensors -- that is, of the CDD field variables formulated by \citet{Hochrainer2015_PhilMag} and also used in the present work. Knowledge of the averaged energy functional together with a correct average representation of the kinematics means that a consistent dynamic theory can be derived without any need for making arbitrary constitutive assumptions. The present paper aims at finalizing the kinematic side of this project.  

\section{Theory}
\label{sec:theory}

\subsection{Mathematical notations and conventions}
In the following $\times$ denotes the cross product between two vectors, a dot '$\cdot$' denotes a single contraction and $\otimes$ is the tensor product. We consider dislocations moving on a single slip system with Burgers vector $\Bb$ and slip plane normal vector $\Bn$. Calculations are performed using a Cartesian coordinate system with basis vectors $\Be_1,\Be_2,\Be_3$ where without loss of generality we set $\Be_1= \Bb/|\Bb|$, $\Be_2= \Ba/|\Ba|$ with  $\Ba= \Bn\times\Bb$ and $\Be_3= \Bn$. We consider dislocations moving by glide only, hence the dislocation lines are contained in parallel glide planes. The spatial direction of the dislocation line can thus be characterized by a single angle $\varphi$, which we take to be the angle between the line tangent vector $\Bl$ and the Burgers vector such that $\Bl(\varphi)=\cos(\varphi)\Be_1 + \sin(\varphi)\Be_2$. If $\Bl(\varphi)$ points into the direction of motion as we move counter-clockwise around a dislocation loop, the loop is termed positive, otherwise it is termed negative. By convention, positive loops expand and negative loops shrink under positive resolved shear stress. 

The operator $()^{\perp}$ defined by $\Bt^{\perp}= \Bt \times \Bn$ rotates a vector $\Bt$ that is contained in the slip plane by $90^\circ$ around the slip plane normal. This operator can alternatively be expressed as $\Bt^{\perp}= \Bt.\Bve$ where $\Bve$ is the second order Levi-Civita tensor in the slip plane coordinate system, with components $\varepsilon^{ij}=\varepsilon^{ijk}n_k$ where $n_k$ are the components of $\Bn$ (we use the Einstein summation convention throughout). A partial derivative w.r.t. $x_i$ is denoted by $\partial_i\equiv \frac{\partial}{\partial x_i}$. The nabla operator of spatial derivatives is defined as the vector operator $\nabla$ = $\Be_i \partial_i$, the curl operator is $\curl=\nabla\times$. 

$\Bt^{\otimes m}$ is the  $m^{\text{th}}$ order completely symmetric tensor obtained from the vector $\Bt$ according to the recursion relation $\Bt^{\otimes 1} =\Bt$, $\Bt^{\otimes m+1} =\Bt^{\otimes m}\otimes\Bt$. $\BA^{(m)}$ is the  $m^{\text{th}}$ order symmetric alignment tensor of an orientation distribution function $\Phi(\varphi)$ and defined as  $\BA^{(m)}=\oint \, \Phi(\varphi) \Bl(\varphi)^{\otimes m} \text{d}\varphi$ . $\text{Tr}(\BT)$ gives the trace of a symmetric tensor $\BT$ through summation over any two indices of $\BT$. The operator $[\BT]_{\text {sym}}$ completely symmetrizes a tensor by averaging its components over all possible permutations of indices.

\subsection{The higher dimensional continuum theory of dislocations}
\label{sec:theory:hdCDD}

Here, we only present a brief summary of the higher dimensional continuum theory of dislocations; for details the reader is referred to the published literature 
\citep{Hochrainer2007_PhilMag,Sandfeld2010_PhilMag90,Hochrainer2014JMPS}. Dislocation lines are envisaged as lifted curves in a configuration space where each point $(\Br,\varphi)$ contains the spatial  point $\Br$ and the orientation angle $\varphi$. To describe the kinematics of single lines within this space we introduce a generalized line direction $\BL$ and a generalized velocity $\BV$, which denote the tangent to the lifted line and the velocity of the lifted line, respectively: 
\begin{align}
\BL\rphi &= (\cos\vp,\sin\vp,k\rphi)=(\Bl(\vp),k\rphi)\\
\BV\rphi &= ( v\sin\vp, -v\cos\vp, \vt\rphi). \label{eq:V}
\end{align}
Here $\Bl$ is the spatial line direction, $v$ is the modulus of the spatial velocity $\Bv$ (the first two terms in \eqref{eq:V}) which is normal to $\Bl$, and $k$ is the line curvature. The additional component of the lifted velocity vector, $\vt\rphi=-\nabla_\BL v\rphi$, is a rotational velocity which causes a line to move in $\vp$ direction, i.e. to change its orientation. $\nabla_\BL$ denotes the gradient along the generalized line direction.
With the above notations it is possible to define the so-called 'dislocation density tensor of second order' $\alphaII$ along with its evolution equations. This tensor can be expressed in terms of a density function $\rho\rphi$ which gives the density of dislocations for each line orientation separately as 
\begin{equation}
\alphaII\rphi=\rho\rphi\BL\rphi\otimes \Bb.
\end{equation}
%
The generalized line direction $\BL$ contains, besides the spatial line direction $\Bl$, as additional component the local curvature $k$. Accordingly, the corresponding components of the higher order dislocation density tensor contain a new variable, the product of the orientation-dependent dislocation density and the local curvature, which we denote as curvature density $q\rphi=\rho\rphi\, k\rphi$. In analogy to the condition that $\Balpha$ is divergence free, one can show that also $\alphaII$ is solenoidal, i.e.,
\begin{equation} \label{eq:solenoidality}
\Div\,\alphaII=0 \;\Leftrightarrow\; \cos\vp\partial_x(\rho)+\sin\vp\partial_y(\rho)+\partial_\vp(q) =0,
\end{equation} 
 where $\Div(\bullet) := \partial_x(\bullet)+\partial_y(\bullet)+\partial_\vp(\bullet)$. \Eqref{eq:solenoidality} reflects the physical fact that dislocation lines do not start or end inside a crystal. The evolution equation for $\alphaII$ is
\begin{align}
\partial_t\alphaII\rphi=-\curl(\BV\rphi\times\alphaII\rphi ).
\end{align}
This evolution equation for $\alphaII\rphi$ can be re-written in terms of two evolution equations for the scalar dislocation density $\rho\rphi$ and the curvature density $q\rphi$:
\begin{align}
\partial_t\rho &=-\Div(\rho \BV)+ qv \label{eq:drhodt}\\
\partial_t q &=  -\Div(q \BV -\rho\BL\vt)\label{eq:dqdt}
\end{align}
 In \eqref{eq:drhodt}, the first term governs the transport of scalar density in the configuration space, while the last term is a source term and accounts for the change of density due to the expansion or shrinkage of loops. Similarly, the terms in \eqref{eq:dqdt} govern the transport of the scalar curvature density in the configuration space.

\subsection{Reducing the higher dimensional continuum theory} \label{subsec:simplfying_HdCDD}

If all dislocations contained in a volume element share the same orientation, and thus move in the same direction, the extension to a higher dimensional configuration space which is at the core of hdCDD is redundant. However, the advantages of Hochrainer's formulation become manifest as soon as we consider averaged dislocation fields where dislocations of different orientation are contained within the same volume element. Since the differently oriented segments inhabit different locations in configuration space, their motion in different directions can be resolved without problems. However, this capability comes at a very substantial price: First, the set of evolution equations, \eqref{eq:drhodt} and \eqref{eq:dqdt} are from a numerical point of view difficult so solve due to the coupling of spatial and orientation degrees of freedom. Second, the additional orientation dimension introduces a large number of additional degrees of freedom, because this direction needs to be discretised such that gradients along the orientation direction can be properly resolved. This led to efforts towards simplifying the higher-dimensional evolution equations while preserving their performance in capturing dislocation kinematics \citep{Hochrainer2009_ICNAAM,Sandfeld_JMR,Hochrainer2014JMPS}. Such simplifications can be constructed in a systematic manner by performing a Fourier expansion of the functions $\rho\rphi$ and $q\rphi$ and using the resulting  Fourier  coefficients as variables of a simplified continuum theory (CDD). Recently \citet{Hochrainer2015_PhilMag} revisited this problem and introduced a mathematically more concise and elegant formulation which uses an expansion of the dislocation density and curvature density function in terms of \textit{symmetric alignment tensors}. In the following, we just state the final results. Assuming pure glide motion for dislocation lines, the reducible symmetric alignment tensors of the dislocation and curvature densities are defined as:  
\begin{align}
\label{eq:An}
\Brho^{(n)}(\Br) &\defeq \oint \,\!\!\rho\rphi 
{\Bl(\vp)}^{ \otimes n}
\,\text{d}\vp. \\
\label{eq:qn}
\Bq^{(n)}(\Br) &\defeq \oint \,\!\! q\rphi\, [\Bve \cdot \Bl(\vp) \otimes
\Bl(\vp)^{ \otimes n-1}]^{\text {s}}
\,\text{d}\vp. 
\end{align}
The symmetric alignment tensors are density-like quantities which characterize the dislocation microstructure. The zeroth order alignment tensors of $\rho\rphi$ and $q\rphi$ are:
\begin{align} \label{eq:rhot}
	\rhot(\Br):=\AO(\Br):=\oint \,\!\! \rho\rphi \,\text{d}\vp\quad \text{and}\quad 
	\qt(\Br):=\Bq^{(0)}(\Br):=\oint \,\!\! q\rphi \,\text{d}\vp.
\end{align}
Their physical meaning is as follows: $\rhot$ gives total dislocation line length per unit volume, irrespective of orientation. The zeroth-order alignment tensor $\qt$ is the line length per unit volume, divided by the line curvature radius. If averaged over a volume of sufficient size, this quantity can alternatively be envisaged as a volume density of dislocation loops.

The first order dislocation density alignment tensor is a vector which measures the excess (geometrically necessary) dislocation density :
\begin{align}\label{eq:Bkappa}
	\AI(\Br)=\oint \,\!\!\rho\rphi\Bl(\vp) \,\text{d}\vp 
\end{align}
The components of this dislocation density vector $\AI(\Br)=[\rho_1^{(1)}(\Br), \rho_2^{(1)}(\Br)]$ are the screw and edge components of the GND density. \AI relates to the curl of the plastic distortion (arising from slip on the single considered slip system), and hence to the classical dislocation density tensor, by:
\begin{align}\label{eq:_Bkappa_alpha}
\AI \otimes \Bb &= \Balpha =- \curl \Bbeta^\text{pl}
\end{align}

The second-order dislocation density alignment tensor is given by
\begin{align}\label{eq:AII}
	\AII(\Br)=\oint \,\!\!\rho\rphi\Bl(\vp)\otimes\Bl(\vp) \,\text{d}\vp.
\end{align}
The components $\AIIoneone$ and $\AIItwotwo$ of this tensor can be understood as total densities of screw and edge components of the dislocation lines. Their sum, Tr$(\AII)$, is equal to $\rhot$. This is an example of the general relation 
\begin{align}\label{eq:rho-trace}
\text{Tr}\, \Brho^{(n)}=\Brho^{(n-2)}\quad \text{and}\quad \text{Tr}\, \Bq^{(n)}=\frac{n-2}{n}\Bq^{(n-2)},
\end{align}

Higher-order dislocation density alignment tensors are less straightforward to interpret in physical terms. However, we think that the above defined quantities provide a sufficiently accurate statistical characterization of dislocation microstructures that accounts for all physically relevant distinctions: The distinction between screw and edge orientations (relevant because of potentially different physical properties such as mobility and cross slip) and the distinction between dislocations associated with strain gradients (the dislocation density vector which relates to the curl of the plastic distortion, or equivalently to the classical dislocation density tensor) and dislocations which are not. In particular, a theory using only the above quantities is perfectly capable of handling situations where locally the dislocation density tensor (the averaged curl of the plastic distortion) is zero but the total, screw and edge dislocation densities are not.

The evolution equations for $\Brho^{(n)}$ and $\Bq^{(n)}$ form an infinite hierarchy where the evolution of lower-order tensors depends on higher order tensors. In the case of an isotropic dislocation velocity they are given by
\begin{align}
\label{eq:dA0dt}
\partial_t\AO &=\nabla \cdot(v \Bve\cdot\AI )+v\BQ^{(0)} \\
\label{eq:dAkdt}
\partial_t\Brho^{(n)} &=\left[\nabla \cdot(v \Bve\otimes\Brho^{(n-1)} )+(n-1)v\BQ^{(n)}-(n-1) \Bve\cdot\Brho^{(n+1)}\cdot\nabla v \right]_{\text{sym}}\\
\label{eq:dqtdt}
\partial_t\Bq^{(0)} &=\nabla \cdot(v\BQ^{(1)} -  \Brho^{(2)}\cdot\nabla v ) \\
\label{eq:dqkdt}
\partial_t\Bq^{(n)} &=\left[\nabla \cdot(v\BQ^{(n+1)} - \Bve\cdot \Brho^{(n+2)}\cdot\nabla v ) \right]_{\text{sym}}
\end{align}
where $\BQ^{(n)}$ are auxiliary symmetric curvature tensors defined as 
\begin{align}
\label{eq:Qn}
\BQ^{(n)}(\Br) &\defeq \oint \,\!\! q\rphi\ \Bve \cdot \Bl{(\vp)} \otimes\Bve \cdot \Bl{(\vp)} \otimes
\Bl^{ \otimes n-2}
\,\text{d}\vp. 
\end{align}
In order to close the hierarchy of evolution equations for the dislocation density alignment tensors $\Brho^{(n)}$ at order $n$ we need to express $\Brho^{(n+1)}$ and $\BQ^{(n)}$ in terms of lower order tensors. We note that a matching closure relation for the curvature density alignment tensors $\Bq^{(n)}$ is not required, because for $n \ge 1$ they can be derived from those for $\rho$ using the solenoidality condition \eqref{eq:solenoidality}. This yields
\begin{align}
	\label{eq:Q}
	\Bq^{(n)}(\Br) &= \frac{1}{n} \left[\div \Brho^{(n+1)}(\Br)\right]_{\text{sym}}.
	\end{align}
The only variable which cannot be expressed in this manner is the zeroth-order alignment tensor $\Bq^{(0)}=\qt$, which must be considered as an independent dynamic variable of the theory. 

\subsection{Recovering the dislocation orientation distribution function (DODF) based on the Maximum Information Entropy Principle (MIEP)}
\label{subsec:DODF}
In \secref{subsec:simplfying_HdCDD} we concluded that the evolution equations of the dislocation density function $\rho\rphi$ can be expressed in terms of its alignment tensors. The information about dislocation orientations is contained in the ``dislocation orientation distribution function" (DODF)  which is defined as $p_{\Br}({\vp})=\rho\rphi/\rhot$ and gives the probability density for dislocations to have, in a given spatial point, a certain orientation $\varphi$. In the following we shall drop, for simplicity of notation, the explicit dependency on $\Br$ with the understanding that $p$ must be considered in each spatial point separately. 

The infinite set of alignment tensors contains complete information about the orientation distribution function, and conversely, closing the hierarchy at any given order $n$ results in a loss of information: Instead of a single orientation distribution, we now have an infinite number of possible distribution functions which are consistent with the known low-order alignment tensors. To estimate the full distribution function based on the limited information contained in the known alignment tensors, one may use the Maximum Information Entropy Principle (MIEP) which states that among all possible distributions the most probable is the one that maximizes the associated Shannon information entropy defined as:
\begin{align}
\label{eq:H}
H \defeq - \oint \, p(\vp) \ln(p(\vp)) d\vp,
\end{align}
The "best guess" for $p(\vp)$ obtained by maximizing $H$ needs to be consistent with the information contained in the $n$ known dislocations alignment tensors, hence it must fulfill the constraints
\begin{align}
\label{eq:constraints_0}
\oint \,\!\!p({\vp})   \,\text{d}\vp &=1 ,
\quad\ldots\quad,\quad
\oint \,\!\!p({\vp}) \Bl^{ \otimes n}  \,\text{d}\vp =\Brho^{(n)}/\rhot.
\end{align}
These constraints are however not independent since, from the alignment tensors of order $n$ and $n-1$, all lower-order alignment tensors can be constructed according to Eq. \eqref{eq:rho-trace}. Thus, we are only dealing with $2n+1$ independent constraints. For these we could consider the $2n+1$ independent components
of the completely symmetric tensors $\Brho^{(n)}$ and $\Brho^{(n-1)}$. However, for our later considerations it is more convenient to use an alternative choice and 
define $2n+1$ independent constraints in terms of the first two components $\rho^{(k)}_{11\dots1}$ and $\rho^{(k)}_{21\dots1}$ of the alignment tensors at each order $k \le n$: 
\begin{align}
\label{eq:constraints_11}
&a)\; n+1\;\text{constraints, \, for all}\; 0\leq k \leq n:\qquad &&\oint \,\!\!p({\vp}) \cos^k   \,\text{d}\vp =\rho^{(k)}_{11\dots1}/\rhot \\
\label{eq:constraints_21}
&b)\; n\;\text{constraints, \, for all}\; 1 \leq k \leq n: \qquad &&\oint \,\!\!p({\vp}) \sin(\vp) \cos^{k-1} (\vp)   \,\text{d}\vp =\rho^{(k)}_{21\dots1}/\rhot.
\end{align}
We now use the maximum entropy method for deriving a 'best guess' for the DODF that is consistent with this set of constraints. To this end, we maximize the entropy and use the standard method of Lagrangian multipliers to account for the constraints. With Lagrangian multipliers $\lambda_i$ and $\lambda'_i$ corresponding to the constraints given by Eqs. \eqref{eq:constraints_11} and \eqref{eq:constraints_21}, respectively, we obtain
\begin{align}
\label{eq:LMM1}
\delta \oint \, \!\!\left( 
p({\vp})\ln(p({\vp})) -\lambda_0  p({\vp}) - \sum_{i=1}^{n} \lambda_i  p({\vp}) \cos^{n}(\vp) 	
-\sum_{i=1}^{n} \lambda'_i  p({\vp}) \cos^{n-1}(\vp)\sin(\vp)	\right)\,\text{d}\vp = 0\\
%
\label{eq:LMM2}
\Rightarrow p({\vp})\left( 
\ln(p({\vp})) +1-\lambda_0 - \sum_{i=1}^{n} \lambda_i   \cos^{n}(\vp) 	
-\sum_{i=1}^{n} \lambda'_i \ \sin(\vp)\cos^{n-1}(\vp)	\right)\,\delta\vp =0.
\end{align}
Since this must be zero for any $\delta\vp$, the terms inside the parentheses must add to zero, which gives 
\begin{align}
\label{eq:DODF}
p({\vp})&=\exp\left(-\mu-\sum_{i=1}^{n} \lambda_i \cos^i(\vp) -\sum_{i=1}^{n} \lambda'_i  \cos^{n-1}(\vp)\sin(\vp)  \right).
\end{align}
The $\lambda_i$ and $\lambda'_i$   are found when the constraints are satisfied. The first constraint gives: 
\begin{align}
1=\oint \,\!\!p({\vp}) \,\text{d}\vp = e^{-\mu}\oint \,\!\! \exp\left(-\sum_{i=1}^{n} \lambda_i \cos^i ({\vp}) + \lambda'_i  \cos^{n-1}(\vp)\sin(\vp)  \right) 
\end{align}
The integral is called the \textit{partition function} of the distribution, 
\begin{align}
\label{eq:z}
Z(\lambda_1,\dots,\lambda_n)&\defeq \oint \,\!\! \exp\left(-\sum_{i=1}^{n} \lambda_i \cos^i(\vp)+ \lambda'_i  \cos^{n-1}(\vp)\sin(\vp) \right).
\end{align} 
Hence
\begin{align}
\label{eq:mu}
\mu&= \ln(Z(\lambda_1,\dots,\lambda_n))
\end{align} 
The remaining series of constraints are satisfied by
\begin{align}\label{eq:lambda_i1}
\frac{\rho^{(k)}_{11\dots1}}{\rhot}&=\oint \,\!\!p({\vp}) \cos^i ({\vp})  \,\text{d}\vp = e^{-\mu}\oint \,\!\! \cos^i ({\vp})\exp\left(-\sum_{i=1}^{n} \lambda_i \cos^i(\vp)  + \lambda'_i  \cos^{n-1}(\vp)\sin(\vp)\right) \nonumber\\
&=\frac{1}{Z}\frac{\partial Z(\lambda_1,\lambda'_1,\dots,\lambda_n,\lambda'_n)}{\partial\lambda_i}=-\frac{\partial \ln Z}{\partial\lambda_i}\\
\label{eq:lambda_i2}
\frac{\rho^{(k)}_{21\dots1}}{\rhot}&=\oint \,\!\!p({\vp}) \cos^i ({\vp})  \,\text{d}\vp = e^{-\mu}\oint \,\!\! \cos^i ({\vp})\exp\left(-\sum_{i=1}^{n} \lambda_i \cos^i(\vp) + \lambda'_i  \cos^{n-1}(\vp)\sin(\vp) \right) \nonumber\\
&=\frac{1}{Z}\frac{\partial Z(\lambda_1,\lambda'_1,\dots,\lambda_n,\lambda'_n)}{\partial\lambda'_i}=-\frac{\partial \ln Z}{\partial\lambda'_i}
\end{align}
which gives a system of $2n$ implicit equations. Solving this system of equations we can find the $2n$ unknown Lagrangian multipliers $\lambda_i$ and $\lambda'_i$ and determine the corresponding maximum-entropy estimate of the DODF:
\begin{align}
\label{eq:DODF2}
p({\vp})&= \frac{1}{Z}\exp\left(-\sum_{i=1}^{n} \lambda_i \cos^i(\vp) + \lambda'_i  \cos^{n-1}(\vp)\sin(\vp)  \right) 
\end{align}
To understand the relevance of this function we need to discuss it from an information-theoretical point of view. The more we know about the orientations of dislocations, the smaller is the information actually contained in the variable $\varphi$. If we take the extreme case where all dislocations are straight parallel edges of one sign, the information contained in the variable $\varphi$ is actually zero since determining the orientation of any randomly chosen dislocation would not tell us anything that we do not already know. Maximizing the information under a set of constraints is tantamount to choosing a DODF which accounts for the information contained in the constraints - which reduces the information gained by determining the orientation of a dislocation - {\em but no other information}. Now, if the constraints are the known variables of a dislocation field theory, the maximum entropy principle allows us to construct a DODF that is consistent with the values of these variables but does not imply any knowledge about the orientation distribution of dislocations that is {\em not} contained in the given theory. Conversely, any other choice of DODF would imply that we pretend to know more about dislocation orientations than we can actually learn from our theory. The MIEP as used here provides a systematic method for dealing with the incompleteness of information that is implicit in any simplified, statistically averaged theory: It provides mathematical tools for handling information. The Shannon entropy invoked in this principle is a generic measure of information. Even though in certain cases a thermodynamic interpretation of the Shannon entropy is feasible or even illuminating, we emphasize that in the present case the information entropy associated  with dislocation orientation distributions has little to do with the thermodynamic entropy of the dislocation system. We would, in fact, strongly object to any interpretation that might lead the reader to the conclusion that the properties of dislocation systems might be deduced from a generic thermodynamic principle of entropy maximization.   

In practical terms, we can use the DODF given by Eq. \eqref{eq:DODF2} for two purposes: First, it allows us to provide an approximation to the exact DODF based on a limited number of alignment tensors. We can compare this approximation with the exact DODF that we may derive either from hdCDD or from systematic analysis of DDD simulations, and use this comparison to assess the performance of a given CDD theory. This is illustrated below in \secref{subsec:benchmark_CDD}. Second and more importantly, we can use the maximum-entropy DODF, Eq. \eqref{eq:DODF2}, to evaluate dislocation density alignment tensors of order higher than $n$. Since the parameters of the DODF depend on the alignment tensors of order up to $n$, this provides us with a method of closing the infinite hierarchy of equations, Eq. \eqref{eq:dAkdt}, at any desired order. We note that the resulting closure relations are optimal in the information-theoretical sense: Any other choice of closure relation, such as the ad-hoc relations used e.g. by \citet{Hochrainer2014JMPS,Hochrainer2015_PhilMag}, implies a different form of the DODF with less information, and is therefore tantamount to introducing hidden assumptions about the orientation distribution of dislocations that are not covered by the information contained in a $n$th order CDD theory. 

In order to use the DODF \eqref{eq:DODF2}, we need to evaluate the Lagrangian multipliers $\lambda_i(\Brho^{(k)}_{11\dots1},\Brho^{(k)}_{21\dots1})$, $\lambda'_i(\Brho^{(k)}_{11\dots1},\Brho^{(k)}_{21\dots1})$ in terms of the alignment tensor components. The analytical solution of the system of equations \eqref{eq:lambda_i1} is, except in the case $n=1$ studied by \citet{monavari2014comparison}, not straightforward. Instead we use a numerical approach where we tabulate the functions $\lambda_i(\Brho^{(k)}_{11\dots1},\Brho^{(k)}_{21\dots1})$,
$\lambda'_i(\Brho^{(k)}_{11\dots1},\Brho^{(k)}_{21\dots1})$. Using these tabulated functions then allows us to evaluate alignment tensors of order higher than $n$ from the corresponding DODF,
\begin{align}
\label{eq:DODF_moments_ftens}
\Brho^{(k)}& =\rhot \Bf_k(\lambda_1,\lambda'_1,\dots,\lambda_n,\lambda'_n)=\rhot \Bg_k(\AI,\dots,\Brho^{(n)}).
\end{align}
Closure at first order requires consideration of the case $n=1,k=2$. Closure at order $n=2$ requires to evaluate these functions for $n=2,k=3$. In both cases which are studied explicitly later on we find that the functions $\Bf$ and $\Bg$ can be well approximated in terms of elementary functions which provide us with explicit semi-analytical closure relations. 

\subsection{Symmetric DODF and dislocation moment functions}
\label{subsec:symDODF}

Assuming the DODF to possess an axis of symmetry w.r.t. some angle $\vp_{\Brho}$ can further simplify the closure assumptions. To understand the implications of such an assumption we note that anisotropy of the dislocation orientations may arise from two sources: (i) the dislocation velocity may be anisotropic because of some external constraint - a paradigmatic example being dislocations piling up against an internal boundary; (ii) anisotropy may arise from the fact that edge and screw dislocations may have different properties (line energy, mobility, ability to cross slip....). If (i) is the only source of anisotropy, then the DODF will be symmetric with respect to the angle which marks the orientation of the boundary. If (ii) is the only source of anisotropy, then the DODF will be symmetric with respect to the screw and edge orientations. Both effects may work in conjunction, such as in the piling up of edge dislocations against the walls of a PSB microstructure \citep{Mughrabi1983_Acta} where the DODF must be symmetric with respect to the screw orientation. Only if the 'external' source of anisotropy (i) and the 'internal' source of anisotropy (ii) favor different axes of symmetry, as in the example of a persistent slip band impinging obliquely on a grain boundary, the assumption of a symmetrical DODF is unwarranted. 

\begin{figure}
  		\centering
  		\includegraphics[width=1\linewidth]{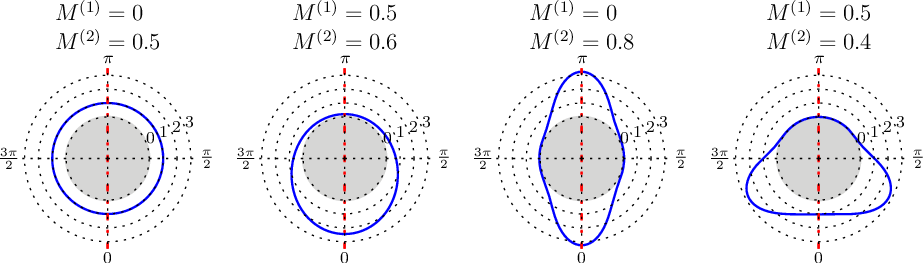}
  		\caption{Symmetric DODFs constructed from the first two moment functions $M^{(1)}$ and $M^{(2)}$; plotted is $2 \pi p(\varphi)$  as function of the angle $\varphi$, $\varphi = 0,\pi$ are screw orientations, $\varphi = \pi/2,2\pi/2$ are edge orientations. From left to right: DODF with isotropically distributed SSD, GND dominated DODF with GND in screw orientation, DODF dominated by screw dislocation dipoles without GND,  and DODF with both geometrically necessary screw dislocations and edge dislocation dipoles.  While  these DODFs are admissible in \CDDII, only the first two DODFs can be represented by \CDDI.    }
  		\label{fig:DODF}
  	\end{figure}

If the DODF is symmetrical, the alignment  tensor of GND density $\AI$ (the dislocation density vector) points along the direction given by the symmetry angle $\varphi_{\Brho}$. The corresponding line direction $\lk$ and its perpendicular $\lkp$ are
s\begin{align}
\lk &= \AI/|\AI| =[\lkone,\lktwo]=[\cos(\vp_{\Brho}),\sin(\vp_{\Brho})]
\qquad\text{and}\qquad
 \lkp =  \lk\Bve=[\sin(\vp_{\Brho}),-\cos(\vp_{\Brho})]. 
\end{align} 
Transforming the coordinates of the orientation space to $\psi=\varphi-\varphi_{\Brho}$ gives:
\begin{align}
\label{eq:An_psi}
	\Brho'^{(n)} &\defeq \rhot \oint \,\!\!p({\psi})
	\Bl{(\psi)}^{ \otimes n}
    \,\text{d}\psi \\
\text{where}\qquad
\label{eq:l}
\Bl_{(\psi)}&=[\cos(\vp-\vp_{\Brho}),\sin(\vp-\vp_{\Brho})]
\end{align} 
The first  components of the $\Brho'^{(n)}$ tensors are related to the \textit{moment functions}:
\begin{align}
	\label{eq:M_n}
	M^{(n)} &= \frac{\rho'^{(n)}_{11\dots1}(\psi)}{\rhot}=\oint \,\!\!p({\psi}) \cos^n(\psi) \,\text{d}\psi=\oint \,\!\!p({\vp}) \cos^n(\vp-\vp_{\Brho})
    \,\text{d}\vp.
\end{align}
All other components of $\Brho'^{(n)}$ are either zero or can be calculated from the $M^{(n)}(\psi)$  series.
Dislocation moment functions characterize the main features of DODF. For example the first moment is the ratio of GND density to total density and the second moment describes the orientation of SSD w.r.t GND density:
\begin{align}
\label{eq:M_1_M_2}
M^{(1)} &= |\AI|/\rhot,\\
M^{(2)}&=\left(\AIIoneone\lkone\lkone+2\AIIonetwo\lkone\lktwo+\AIItwotwo\lktwo\lktwo\right)/\rhot.
\end{align}  
  	\Figref{fig:DODF} depicts the different DODF constructed by the first two moment functions.

  For symmetrical orientation distribution functions, the first $n$ terms of the dislocation moment function series $M^{(n)}$  together carry the same information as the first $n$ terms of the dislocation alignment tensor series  $\Brho^{(n)}$. In fact, both series can be converted into each other: 
\begin{align}
\label{eq:MtoA}
	\rho^{(m+n)}_{\underbrace{\scriptstyle 1\dots 1 }_{m\, \text{times}}\underbrace{\scriptstyle 2\dots 2 }_{n\, \text{times}}} =& \oint \,\!\!\rho\rphi \cos^m(\varphi)\sin^n(\varphi)\,\text{d}\vp \nonumber\\
=&\oint \,\!\!\rho\rphi \cos^m(\psi+\vp_{\Brho})\sin^n(\psi+\vp_{\Brho})\,\text{d}\vp \nonumber\\
=&\oint \,\!\!\rho\rphi (\cos(\psi)\cos(\vp_{\Brho})-\sin(\psi)\sin(\vp_{\Brho}))^m
(\cos(\psi)\sin(\vp_{\Brho})+\sin(\psi)\cos(\vp_{\Brho}))^n\,\text{d}\vp \nonumber\\
=&\oint \,\!\!\rho\rphi \left[\sum^m_{i=0}\binom{m}{i}\cos^{m-i}(\psi)\cos^{m-i}(\vp_{\Brho})\sin^i(\psi)(-\sin(\vp_{\Brho}))^i\nonumber\right]\\
&\times\left[\sum^n_{j=0}\binom{n}{j}\cos^{n-j}(\psi)\sin^{n-j}(\vp_{\Brho})\sin^j(\psi)\cos^j(\vp_{\Brho})\nonumber\right]\,\text{d}\vp\\
=&\oint \,\!\!\rho\rphi \sum^m_{i=0}\sum^n_{j=0}\binom{m}{i}\binom{n}{j}
\cos^{m+n-i-j}(\psi)\sin^{i+j}(\psi)\nonumber\\
&\times\cos^{m-i}(\vp_{\Brho})(-\sin(\vp_{\Brho}))^i\sin^{n-j}(\vp_{\Brho})\cos^j(\vp_{\Brho})\,\text{d}\vp
\end{align}
where $\binom{m}{i}$ denotes the number of possibilities to choose $i$ elements out of a set of $m$.  
After removing those terms which vanish because of symmetry, \eqref{eq:MtoA}  can be reformulated in tensor notation as
\begin{align}
\label{eq:MtoAK}
\Brho^{(k)}/\rhot&=  M^{(k)}
\overbrace{	\lk \otimes \cdots \otimes \lk   
 }^{k \text{ times}} + (M^{(k-2)}-M^{(k)})
\overbrace{	\lkp \otimes \lkp \otimes\cdots \otimes \lk   }^{\large\binom{k}{2} \text{ are } \lkp}\normalsize \\ \nonumber
&+ (M^{(k-4)}-2M^{(k-2)}-M^{(k)}) \overbrace{	\lkp \otimes \lkp \otimes\cdots \otimes \lk   
 }^{\large \binom{k}{4} \text{ are } \lkp}\normalsize  +\dots 
\end{align}
Following the procedure from \secref{subsec:DODF} the symmetric DODF and its moments become:
\begin{align}
\label{eq:DODF_sym}
p({\psi})&= \frac{1}{Z}\exp\left(-\sum_{i=1}^{n} \lambda_i \cos^i(\psi)   \right) \\
\label{eq:DODF_moments}
M^{(k)}& = \oint \, \frac{1}{Z}\exp\left(-\sum_{i=1}^{n} \lambda_i \cos^i(\psi)   \right) \cos^k(\psi) \,\text{d}\psi,
\end{align}
and again the unknown moments are functions of Lagrangian multipliers or  known dislocation moments:
\begin{align}
\label{eq:DODF_moments_f}
M^{(k)}& =f(\lambda_1,\dots,\lambda_n)= g(M^{(1)},\dots,M^{(n)})=h(\AI,\dots,\Brho^{(n)}).
\end{align}

\subsection{Derivation of the lowest order simplified continuum dislocation dynamics theory - \CDDI}
\label{subsec:CDD1}

The lowest order simplified theory (\CDDI ) is constructed by only considering the total dislocation density $\rhot \equiv  \AO$, the dislocation density vector  $\AI$, and the total curvature density $\qt \equiv \Bq^{(0)}$. The evolution equations for $\rhot$, $\AI$, and $\qt$ as derived from \eqref{eq:dA0dt}-\eqref{eq:dqkdt} then take the form
\begin{align}
\label{eq:drhotdt_CDD1}
\partial_t\rhot &=\nabla \cdot(v \Bve\cdot\AI)+v\qt \\
\label{eq:dA1dt_CDD1}
\partial_t\AI &=\nabla \cdot(v \Bve \rhot)\\
\label{eq:dqdt_CDD1}
\partial_t \qt &=\nabla \cdot(v\BQ^{(1)} -  \Brho^{(2)}\cdot\nabla v ) 
\end{align}

For this theory, the closure problem is to \emph{express the quantities $\BQ^{(1)}$ and $\AII$ in terms of $\qt,\rhot$, and $\AI$}. Exact solutions for this problem are available in two limiting cases: (i) in the limit of isotropic dislocation arrangements without geometrically necessary dislocations, the second order alignment tensor is $\AII = \onetwo\rhot \BI^{(2)}$ where $\BI^{(2)}$ is the second rank unit tensor; (ii) in the limit where all dislocations are geometrically necessary, we find that $\AII = |\AI|(\lk \otimes \lk)$. Any suitable closure assumption should satisfy these two limits.
Following the procedure in \secref{subsec:symDODF}, \eqref{eq:DODF} for the case of \CDDI gives the DODF as:
\begin{align}
\label{eq:CDDI_DODF}
p({\psi})&=\exp\left({ -\mu- \lambda_1 \cos(\psi)}\right).
\end{align}
$\mu$ and $\lambda_1$ are the unknown Lagrangian multipliers and are subject to solution of \eqref{eq:lambda_i1}. \Eqref{eq:MtoAK} gives the unknown  alignment tensor $\AII$ in terms of dislocations moment functions $M^{(2)}$ :
\begin{equation}\label{eq:close_a2_max}
\AII = \rhot\left[ M^{(2)} \lk \otimes \lk + (1-M^{(2)} )\lkp \otimes \lkp\right],
\end{equation}
$M^{(1)}$ and $M^{(2)}$ are smooth functions of Lagrangian multiplier $\lambda_1$ and therefore $M^{(2)}$ can be approximated by the known moment  $M^{(1)}$:
\begin{align}\label{eq:M2}
M^{(2)} &\approx [2 + (M^{(1)})^2 + (M^{(1)})^6]/4,  \quad M^{(1)}=|\AI|/\rhot
\end{align}

In the derivation of the lowest order CDD (simplified CDD) in \citep{Hochrainer2014JMPS} instead a linear interpolation between the two limiting cases was used:
 \begin{equation}\label{eq:close_a2_lin}
 \AII\approx \frac{1}{2}\left[ (\rhot +|\AI|) \lk \otimes \lk + (\rhot -|\AI|)\lkp \otimes \lkp\right].
 \end{equation}

Interestingly this assumption resembles the  structure of \eqref{eq:close_a2_max}. In the limits of $\AI =0$ (isotropic dislocation arrangement), or $\AI= \rhot$ (fully polarized dislocation arrangement), both expressions are equivalent and become exact. However, the derivation based on the MIEP provides a nonlinear interpolation which in the regime of small $\AI$ differs significantly from the linear approximation. An obvious drawback of the linear interpolation is that it is non-analytic in the point $\AI = 0$ \citep{Sandfeld2015b_MSMSE}.
A further closure approximation is needed for the curvature density vector.  An expression which covers both limit cases is given by 
\begin{align}
\label{eq:closeq}
\BQ^{(1)} =- (\AI)^{\perp} \frac{\qt}{\rhot}.
\end{align}
This expression corresponds to an \textit{equi-convex} dislocation microstructure where all dislocations (irrespective of orientation) share the same curvature. Alternatively, the curvature vector can be calculated from the approximation for $\AII$ using \eqref{eq:Q}.  \citet{monavari2014comparison} have studied the consequences of these different closure assumptions.

\subsection{Derivation of the second order simplified continuum dislocation dynamics theory , \CDDII}
\label{subsec:CDD2}
\begin{figure}
    	\centering
    	\includegraphics[width=1\linewidth]{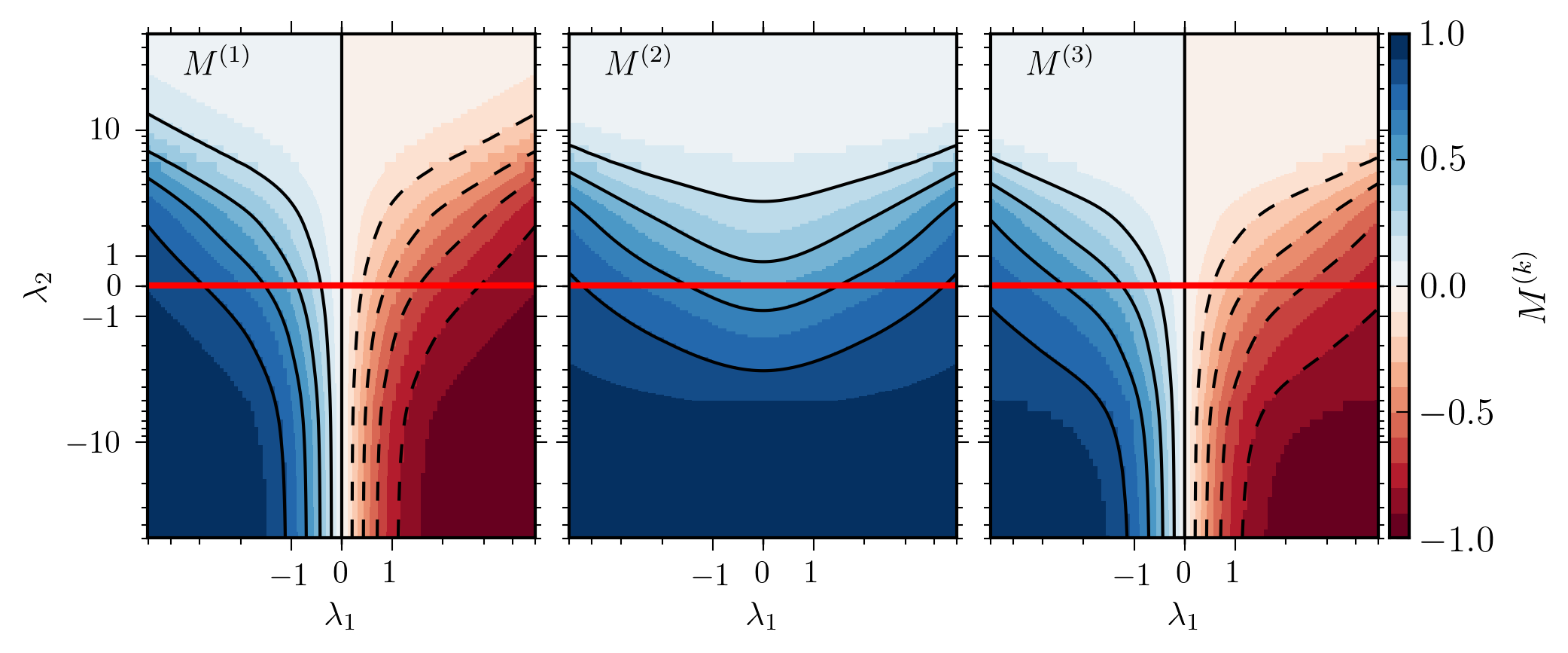}
    	\caption{Visualization of the moment functions  $M^{(1)}\ldots M^{(3)}$ as functions of the two Lagrangian multipliers $\lambda_{1,2}$ of a second-order MIEP orientation distribution function. The red line corresponds to $\lambda_2=0$ which is the subspace of DODFs resulting from \CDDI.}
    	\label{fig:M_l1l2}
    \end{figure}

While the field variables of the lowest order of the CDD equations (\CDDI) carry the information about total dislocation density and line curvature together with geometrically necessary screw and edge densities, the \emph{total} screw and edge dislocation density can only be estimated. The \CDDI\ theory can already represent a wide range of dislocation systems where the dislocations are mostly  geometrically necessary, or where statistically stored dislocation are isotropically distributed over the possible orientations. However, the theory cannot distinguish between statistically stored screw and edge dislocations and must, hence, fail when dislocation arrangements become strongly anisotropic. In order to represent \textit{dipolar} dislocation distributions it is necessary to incorporate the second-order alignment tensor $\AII$ into the formulation.
Closing the evolution equation \eqref{eq:dAkdt} at the second order alignment tensor $\AII$ gives \citep{Hochrainer2015_PhilMag}:
\begin{align}
	\label{eq:dkappadt_A2}
	\partial_t\AI&= \nabla \cdot (v \Bve \otimes \text{Tr}(\AII))\\
	\label{eq:dA2dt}
	\partial_t \AII&=\left[\nabla  \cdot (v \Bve \otimes \AI)+v\QII-\Bve \cdot \AIII\cdot\nabla v \right]_{\textbf{sym}}\\
	\label{eq:dqtdt_A2}
	\partial_t\qt&=\nabla \cdot ( v\BQ^{(1)} - \AII\cdot \nabla v )
\end{align}
Thus, in order to close the theory we need to represent the $\AIII$ and $\QII$ alignment tensors in terms of the lower-order tensors contained in the theory. Following the same procedure as before, we use the MIEP to reconstruct the DODF
\begin{align}
\label{eq:CDDII_DODF}
p({\psi})&=\exp\left({ -\mu- \lambda_1 \cos(\psi)- \lambda_2 \cos^2(\psi)}\right).
\end{align}
and use tabulation to relate the Lagrange multipliers $\lambda_{1,2}$ to the Moment functions $M^{(1,2)}$. 
\Figref{fig:M_l1l2}  shows that the functions $M^{(1)}$, $M^{(2)}$ and  $M^{(3)}$ are smooth non-singular functions of the Lagrangian multipliers. \Figref{fig:DODF} illustrates the shape of the DODF given by \eqref{eq:CDDII_DODF} for different values of the moment functions. 
    \begin{figure}[ht!]
    	\centering
    	\includegraphics[width=.93\linewidth]{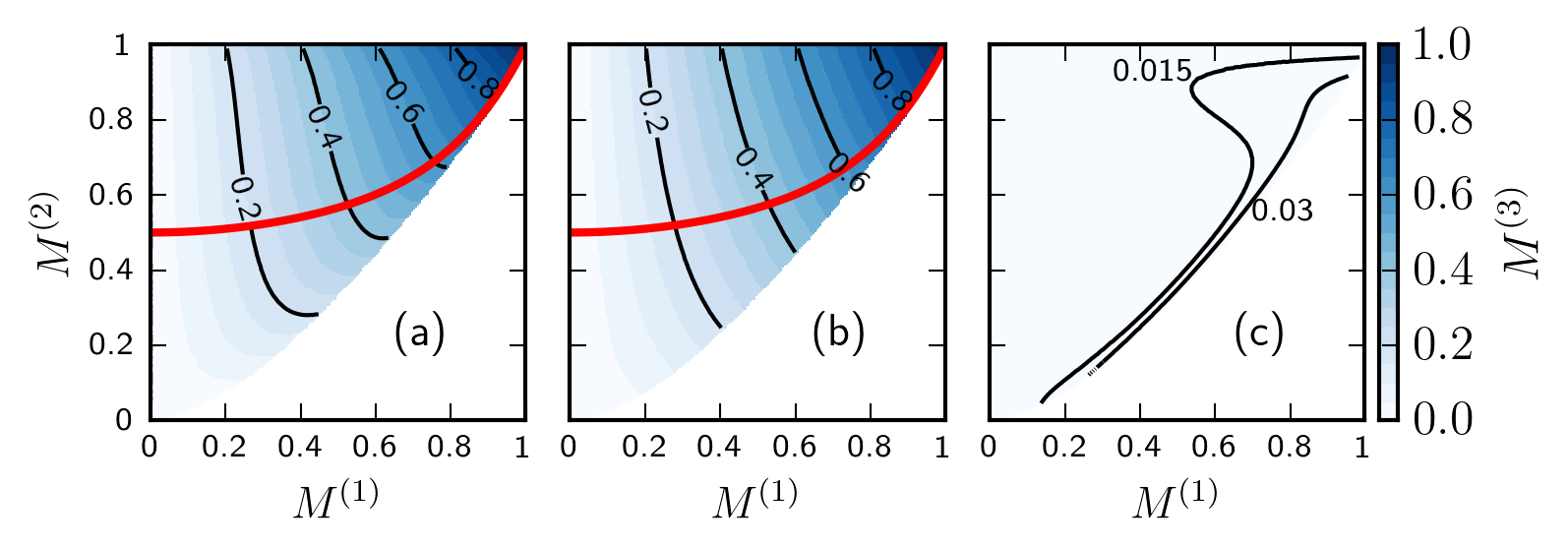}
    	\caption{ Left: $M^{(3)}$ as estimated from $M^{(1)}$ and $M^{(2)}$  using the reconstructed DODF of \CDDI theory. Center: Its analytical approximation: $M^{(3)} \approx M^{(1)}\sqrt{M^{(2)}}$; right graph: Error of the analytical approximation for $M^{(3)}$. Red lines: the admissible $M^{(2)}$ moment in \CDDI. }
    	\label{fig:M3}
    \end{figure}

We can then evaluate the $\AIII$ tensor from Eq. \eqref{eq:MtoAK} as
\begin{align}
\label{eq:A3_symbolic}
\AIII/\rhot=\left[M^{(3)}  \lk\otimes\lk\otimes\lk+\left(M^{(1)}-M^{(3)}\right )(\lk\otimes\lkp\otimes\lkp
+\lkp\otimes\lk\otimes\lkp
+\lkp\otimes\lkp\otimes\lk)\right]
\end{align}
In case of a pure GND microstructure this equation reduces to $\AIII=\AI\otimes\AI\otimes\AI / |\AI|^2  $ which is Hochrainer's ad-hoc assumption for the $\AIII$ tensor. We instead use the full expression where we approximate the MIEP estimate of $M^{(3)}$ as $M^{(3)}\approx M^{(1)}\sqrt{M^{(2)}}$. This approximation is shown in comparison with the MIEP result and the approximation  error in \figref{fig:M3}.  

The second order auxiliary curvature-density tensor $\QII$ can be derived based on $\Bq^{(2)}$ which from \eqref{eq:Q} is given as the divergence of the $\AIII$ tensor\footnote{Note that this analytically exact equation demands a numerical implementation of equations which can provide the gradients terms of $\AIII$. E.g. in our Galerkin finite element framework, which is $C^0$ continuous, this can be achieved by solving for $\AIII$ as additional variables.}:
\begin{align}
\label{eq:QII_equi}
\QII=\frac{1}{2}\qt \BI^{2}-\Bve\cdot\Bq^{(2)} \qquad\text{with}\qquad \Bq^{(2)}=\frac{1}{2}\left[\div \AIII \right]_{\mathbf{sym}}
\end{align}
One can also construct $\QII$ from the information provided by $\qt$ and $\QI$  following the same steps as for the derivation of $\AII$ in the \CDDI\  theory:
\begin{align}
\QII&= \frac{\qt}{2|\Bq^{(1)}|^2}\left[ (1+\Phi) {\Bq^{(1)}} \otimes {\Bq^{(1)}} + (1-\Phi){\Bq^{(1)}}^{\perp} \otimes {\Bq^{(1)}}^{\perp}\right] 
\end{align}
where \cite{Hochrainer2015_PhilMag} makes again a linear assumption by setting $\Phi={|\Bq^{(1)}|}/{\qt}$. Based on the MIEP we obtain instead $\Phi \approx  (|\Bq^{(1)}|/\qt)^2(1+(|\Bq|/\qt)^4)/2$.

\subsection{Numerical implementation}
For the numerical implementation of our CDD models we reformulate the  evolution equations \eqref{eq:dA0dt} and \eqref{eq:dAkdt} and \eqref{eq:dqkdt} in a conservative form which is the most suitable representation of the underlying physics and e.g. has the advantage of numerically preserving the total number of dislocations in the computational domain \citep{Ebrahimi2014}:
\begin{align}
\partial_t\mathbf{u}_i &= \nabla \cdot \mathbf{F}_i(\mathbf{u})+\mathbf{G}_i(\mathbf{u}).
\end{align}
The \lq container variable vector\rq\ $\mathbf{u}$, the vector of test functions  $\mathbf{w}$, the flux matrix  $\mathbf{F}(\mathbf{u})$ and the source vector $\mathbf{G}(\mathbf{u})$ are defined as
\begin{align}
\mathbf{u}=\left[\begin{array}{c}\rhot \\ \Brho^{(n)}  \\\Bq^{(n)} \end{array}\right], \quad
\mathbf{F}(\mathbf{u})=\left[\begin{array}{c}
v \Bve\cdot\AI  \\
v \Bve\otimes\Brho^{(n-1)} \\
v\BQ^{(n+1)} - \Bve\cdot \Brho^{(n+2)}\cdot\nabla v 
\end{array}\right], \quad \\\nonumber
\mathbf{G}(\mathbf{u})=\left[\begin{array}{c}v\qt \\(n-1)v\BQ^{(n)}-(n-1) \Bve\cdot\Brho^{(n+1)}\cdot\nabla v \\0\end{array}\right],
\mathbf{w}=\left[\begin{array}{c} w_\rho^{(0)} \\ w_\Brho^{(n)}  \\w_\Bq^{(n)} \end{array}\right], \quad
\end{align}
For numerically solving this system of equations we use a semi-discontinuous finite element method. In the finite element space $V_h$ the weak formulation of the CDD conservation law is obtained from integration by parts against the vector of  suitable test functions $\mathbf{w}$ such that
\begin{align}
(\partial_t \mathbf{u}_i, \mathbf{w}_i)_{\Omega} &= 
(\nabla \cdot \mathbf{F}_i(\mathbf{u}), \mathbf{w}_i)_{\Omega}+( \mathbf{G}_i(\mathbf{u}), \mathbf{w}_i) + (\mathbf{H}_i(\mathbf{u}^+, \mathbf{u}^-, \mathbf{n}), \mathbf{w}_i^+)_{\partial \Omega}
\\ &\approx ( \mathbf{F}_i(\mathbf{u}), \nabla \cdot \mathbf{w}_i)_{\Omega} + ( \mathbf{G}_i(\mathbf{u}), \mathbf{w}_i) +  (\mathbf{H}_i(\mathbf{u}^+, \mathbf{u}^-, \mathbf{n}), \mathbf{w}_i^+)_{\partial \Omega} + (c\nabla \mathbf{u}, \nabla \mathbf{w}),
\end{align}
where $(a, b)_\Omega=\int_{\Omega} a\, b$. We impose a periodic flux boundary condition using the Lax-Friedrich numerical flux $ \mathbf{H}(\mathbf{u}^+,\mathbf{u}^-,\mathbf{n}) = \frac{1}{2}(\mathbf{F} (\mathbf{u}^+) \cdot \mathbf{n} + \mathbf{F}(\mathbf{u}^-)\cdot \mathbf{n} + \alpha (\mathbf{u}^+ - \mathbf{u}^-))$.
The superscripts '+' and '-' denote the inner and outer traces of a function w.r.t. a finite element cell. The diffusive term $c\Delta \mathbf{u}$ is introduced solely for purposes of numerical stability. The explicit third order Runge-Kutta method is used for time integration which benefits from both low memory usage and high stability properties. This system of partial differential equations is implemented using the  \emph{Differential Equations Analysis Library} (deal.II) \citep{dealII82}. In the following benchmarking exercise where we also consider alternative models proposed in the literature, the same numerical implementation was -- with small adjustments -- used for all considered continuum  models. This guarantees that deviations in the results are not due to different strategies for numerical implementation.

\section{Numerical validation and benchmark test}\label{sec:numerical}
For numerical validation of our kinematic closure procedure we assess the ability of hdCDD, \CDDI\ and \CDDII\ to represent the evolution of dislocation microstructure in a given velocity field, using discrete dislocation dynamics simulation  as a reference. We also benchmark the performance of our equations against the continuum models proposed by \citet{Groma1997_PhysRevB_p5807} and \citet{Arsenlis2004_JMPS52} as well as by \citet{Acharya2006_JMPS} and \citet{Fressengeas2005}. 

A good benchmark test should allow to interpret as many details of the microstructure evolution as possible in physical terms, and it should also allow to demonstrate where the limitations of one or the other model are. At the same time, it is desirable to only test a limited number of model aspects, in order to avoid a large numbers of parameters and interdependencies. We design the benchmark such that a suitable theory needs to be able to 
\begin{itemize}
	\item differentiate between the kinematics of loops and straight lines,
	\item handle the mutual conversion of SSDs and GNDs; we consider this as an essential pre-requisite for any averaged theory that operates on scales above the resolution of individual dislocation lines,	
	\item differentiate between SSDs of screw or edge orientation. 
\end{itemize}
An ideal benchmark situation in this respect is the motion of mobile dislocations in a persistent slip band (PSB) microstructure \figref{fig:PSB_Mughrabi}), for which abundant experimental evidence is available (see e.g. \citet{Mughrabi1983_Acta}) including in-situ observations of dislocation motion (see e.g. \citet{Lepinoux1985_PhilMag}), such that the simulation results can be directly related to experimental information.

\subsection{Formulation of the benchmark problem}
\label{sec:benchmark_system}
\begin{figure}[ht!]
   	\centering
   	\begin{subfigure}[b]{0.45\textwidth}
   		\centering
   		\includegraphics[width=\textwidth]{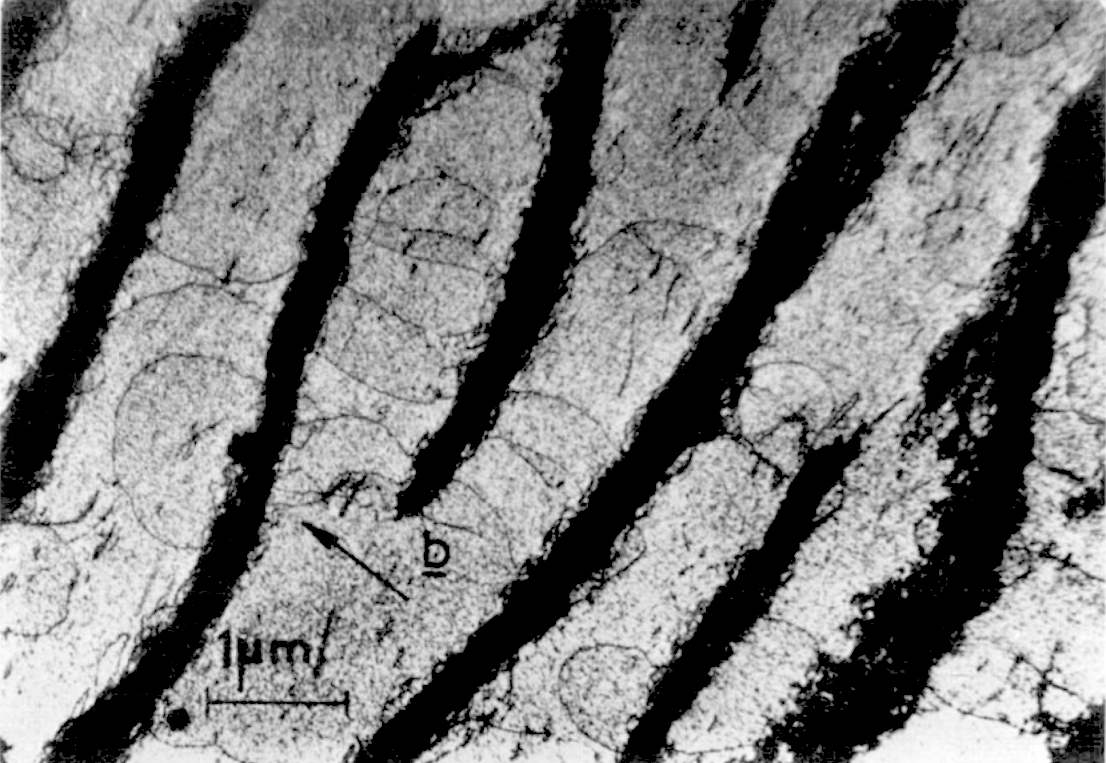}
   		\caption{\small A TEM image of PSBs in copper \citep{Mughrabi1979}.}    
   		\label{fig:PSB_Mughrabi}
   	\end{subfigure}
   	\hfill
   	\begin{subfigure}[b]{0.45\textwidth}  
   		\centering 
   		\includegraphics[width=\textwidth]{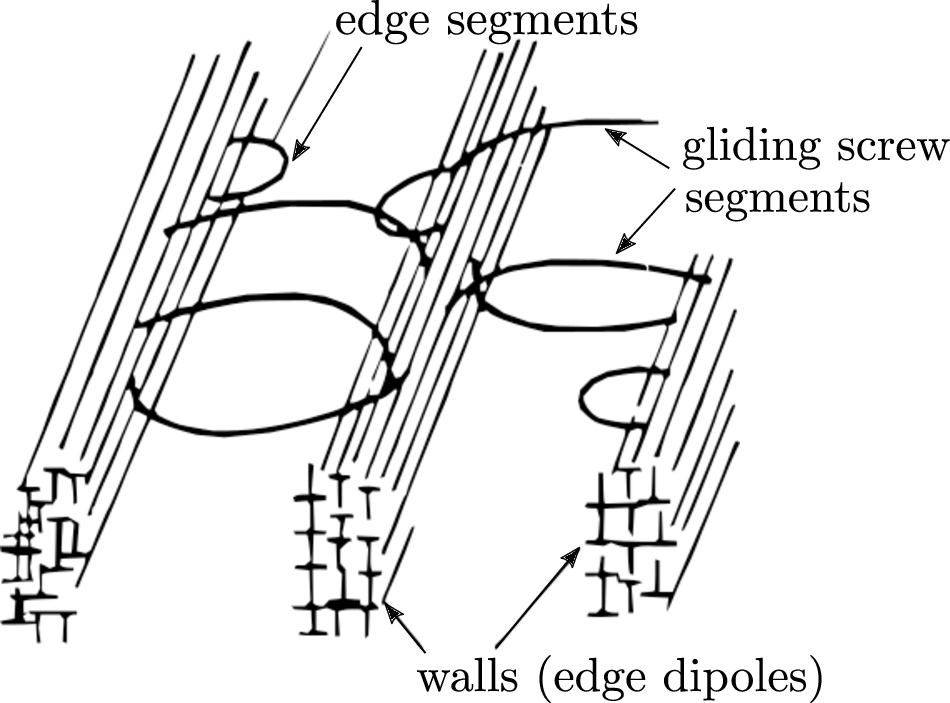}
   		\caption{\small Structure of an idealized PSB reproduced from  \citet{Differt1993_MSEA}}    
   		\label{fig:PSB_Differt}
   	\end{subfigure}
   	\begin{subfigure}[b]{0.45\textwidth}   
   		\centering 
   		\includegraphics[width=\textwidth]{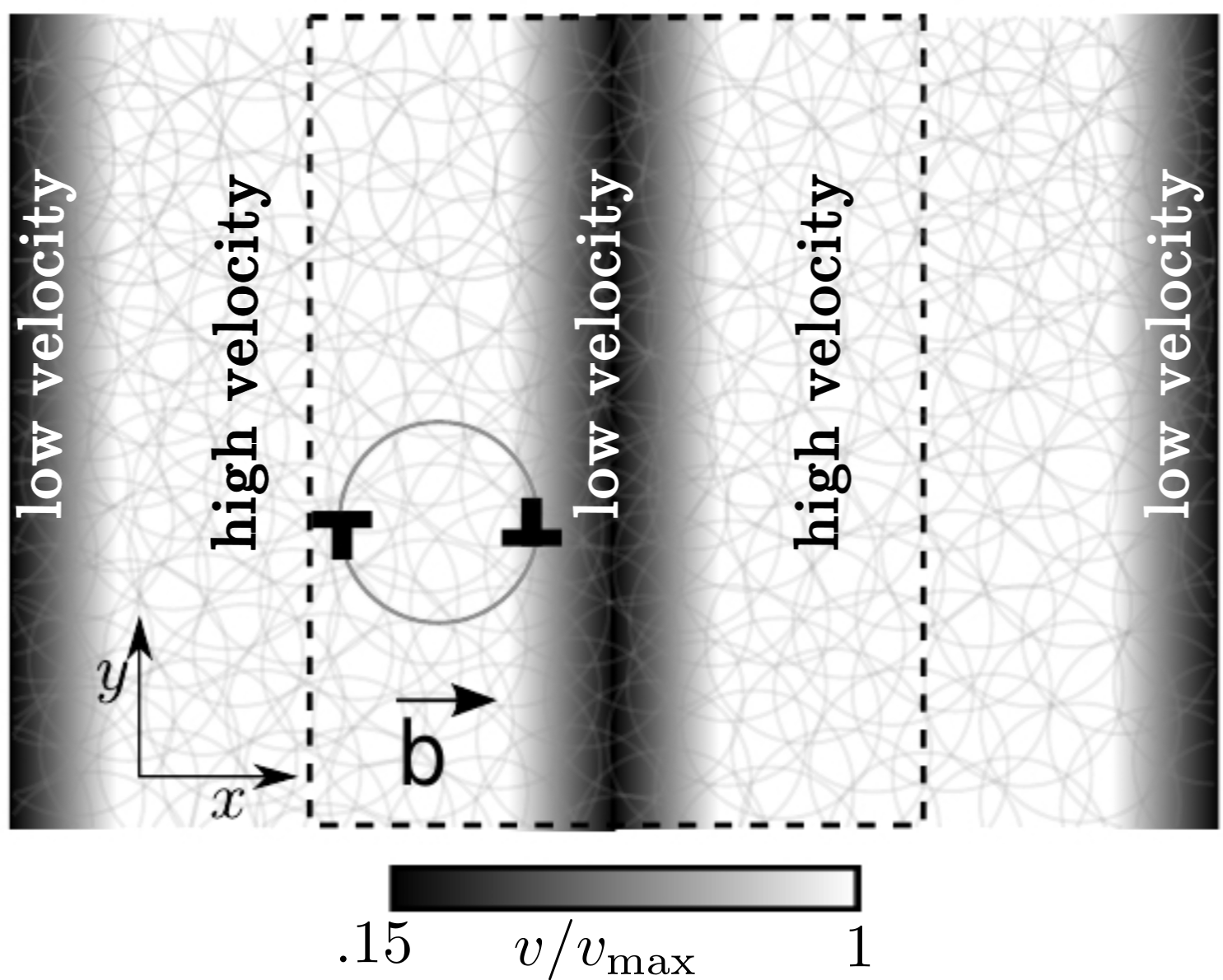}
   		\caption[]%
   		{\small Idealized initial values: prescribed velocity field and uniformly distributed loops}    
   		\label{fig:PSB_IV}
   	\end{subfigure}
   	\quad
   	\begin{subfigure}[b]{0.45\textwidth}   
   		\centering 
   		\includegraphics[width=\textwidth]{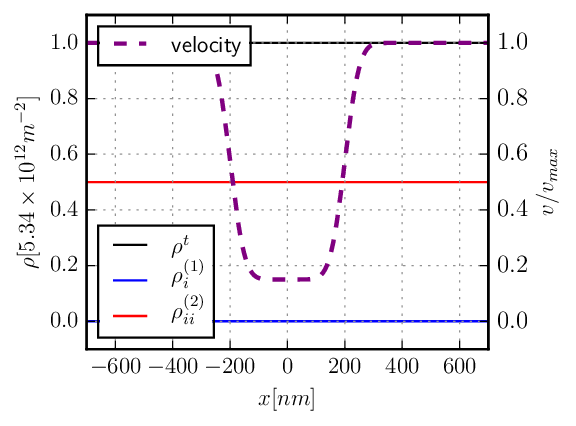}
   		\caption[]%
   		{\small Profile of initial values of dislocation density and velocity in 1.5D simulation box}    
   		\label{fig:PSB_T0}
   	\end{subfigure}
   	\caption{(a) Persistent slip band structure: Low density channels are separated by high density walls which mostly consist of edge dislocations in dipolar formation. Note that dislocations with the same orientation have the same curvature. (b) Sketch of an idealized PSB: dislocations of edge character get trap and bow-out from  walls and screw segments can easily glide through channels . (c) The kinematic setup of idealized PSB: Prescribed velocity is maximum in the channels and drops to 15\% in the middle of the wall (darker shaded area); the initial mobile dislocation population consists of uniformly distributed glide loops; the dashed line indicates the periodic simulation box. (d) Initial dislocations segments are distributed uniformly in all direction which means total dislocation density is consisting of equal amount of edge and screw components.   } 
   	\label{fig:PSB}
   \end{figure}

As benchmark system we consider a PSB structure consisting of a periodic arrangement of obstacle-rich walls separated by obstacle-free 'channels' (see \figref{fig:PSB_Differt}). The 'walls' mainly consist of immobile edge dipoles which hinder the motion of dislocations but do not themselves participate in the plastic deformation process, while in the 'channels' dislocations can move freely. Note that we do not consider the \emph{formation} of such a persistent structure during cyclic loading (for an overview of theoretical models for PSB patterning the reader may consult e.g. \citet{Kubin2002}). Here, we only pursue the modest aim of modeling how dislocations move in such a strongly heterogeneous microstructure by investigating the motion of dislocations over half a loading cycle, in a situation where the obstacle microstructure has already reached a quasi-stationary state - which may persist over thousands of cycles. We describe dislocation-obstacle interactions in terms of a yield stress which depends on the density of immobile edge dislocation dipoles in the walls, $\tau^{\text y} = a \mu b \sqrt{\rho^{\text d,w}}$ where $\mu$ is the shear modulus, $b$ the Burgers vector length, and $a \approx 0.2$ a non-dimensional coefficient characteristic of dipole hardening. The yield stress reduces the resolved shear stress $\tau^{\text res}$ such that the velocity is given by $v = (b/B) (\tau^{\text res}-\tau^{\text y})$ where $B$ is the dislocation drag coefficient. In the following we assume parameters typical of PSB microstructures in Cu, viz, $B = 1\times 10^{-4}$Pa s, $b = 0.256$ nm, $\mu = 4\times 10^4$ MPa, $\alpha = 0.2$, and $\rho^{\text d,w} = 6.2^{14}$ m$^{-2}$, corresponding to a typical stress level for wall deformation of $\approx 60$ MPa. 
The dipole density drops from $\rho^{\text d,w}$ in the walls to zero in the channels, and we model the smooth transition in between by a sigmoidal function. The concomitant velocity profile is shown in \figref{fig:PSB_IV} and assumed time independent. The assumption of a stationary velocity field can be justified by looking at experimental data: The density of immobile dipoles in a fully developed PSB structure ($10^{14}-10^{15}$ m$^{-2}$) is about two orders of magnitude higher than the density of the moving dislocations ($10^{12}-10^{13}$ m$^{-2}$), see \citet{Essmann1979_PhilMag,Mughrabi1983_Acta}. This has two consequences: (i) the dipole density is a slow variable which cannot change much over a half cycle, hence we can assume the obstacle field and the concomitant velocity field as stationary, (ii) the flow stress is controlled by interactions between mobile dislocations and immobile dipoles, rather than among mobile dislocations. Hence we can assume the dislocation velocity to be independent of the densities of the moving dislocations.

In our DDD reference model we consider a simulation volume of size $L_x \times L_y \times L_z = 1.4 \times 15 \times 20\, \mu$m$^3$ with periodic boundary conditions in all three spatial directions. The simulation volume contains $900$ mobile dislocation loops which are initially assumed circular with radius $r_0 = 0.4 \mu$m and which are located at random positions. This configuration corresponds to an initial mobile dislocation density $\rhot = 5.3 \times 10^{12}$  m$^{-2} = \rho_0$. The large values of $L_y$ and $L_z$ are chosen in order to have sufficiently many loops for obtaining meaningful statistical averages.

\subsection{Brief formulation of the benchmark theories}

In the following we benchmark the performance of CDD and of other models proposed in the literature. We consider the model of  \citet{Groma1997_PhysRevB_p5807} for straight parallel dislocations, the edge-screw dynamics of \citet{Arsenlis2004_JMPS52}, and two versions of the semi-phenomenological model of Acharya and co-workers \citep{Acharya2006_JMPS,Fressengeas2005}. 

The problem we consider is statistically homogeneous in the $y$ and $z$ directions and the corresponding simulations can be envisaged as simulations of "1.5D" systems: the computational domain is one-dimensional, dislocation densities depend only on the $x$ coordinate, and any geometrically necessary dislocations must possess edge orientation. However, at the same time the evolution of dislocation densities is influenced by two-dimensional dislocation motions. With the exception of the model of Groma, which considers straight parallel edges only, all considered models contain information about 'perpendicular' motion of dislocations in $y$ direction. Some of the models contain such information in explicit form through variables such as the curvature density in \CDDI\ or \CDDII, or the equations describing screw dislocation motion in the model of Arsenlis et al., Otherwise such information is implicitly contained in parameters describing dislocation multiplication, which is a process that necessarily requires two-dimensional dislocation motion.

\subsubsection{The reference DDD model}
DDD simulations serve as reference providing detailed insight into the discrete microstructure evolution. The used DDD model is based on the so-called Parametric Dislocation Dynamics method (PDD), cf. \citep{Ghoniem:2000td} as implemented in \citet{Po:2014en}. There, cubic splines are used in order to have non-vanishing curvature along each dislocation segment. Since we are here only interested in the kinematic consistency of different models, we evolve the discrete dislocation lines in the same analytically given velocity field as used in the CDD simulations. The evolving discrete dislocation microstructure can be directly compared with the respective continuous fields through the \emph{Discrete-To-Continuum (D2C)} strategy \citep{Sandfeld2015b_MSMSE}: for extracting continuous field data from DDD we start from the discrete line description and convolute them with a non-local smoothing kernel function to obtain continuous and differentiable continuum data that can easily be compared with data obtained from hdCDD/CDD. Further details on the \emph{D2C} strategy can be found in \citep{Sandfeld2015b_MSMSE}. 

\subsubsection{The model of Groma: straight positive and negative edge dislocations}
Groma and co-workers \citep{Groma1997_PhysRevB_p5807,Groma2003_ActaMater} developed a kinematic theory for systems of straight parallel edge dislocations moving in $\pm x$ direction. The microstructure is divided into positive $\rho^+$ and negative $\rho^-$ edge densities (\figref{fig:DODF_GAA}) which  evolve according to
\begin{align}
\label{eq:groma1}
\partial_t\rho^+ =- \partial_{x}\left( \rho^+v\right)
\qquad \text{and} \qquad
\partial_t\rho^- = \partial_{x}\left( \rho^- v\right),
\end{align}
Total density and edge component of GND vector follow from \eqref{eq:groma1}  as
\begin{align}
\label{eq:groma_rhot}
\rhot   = \rho^+ +\rho^-
\qquad\text{and}\qquad
\AItwo   = \rho^+ -\rho^-
\end{align}
To be consistent with our initial conditions, we choose the initial conditions for the Groma model as $\rhot = \rho_0, \AItwo = 0$ corresponding to $\rho^+(x)=\rho^-(x)=\rho_0/2$. 

Note that in the Groma model all dislocations are assumed to be straight, and therefore there can be no dislocation multiplication (the total dislocation line length is constant). We note that various authors have introduced phenomenological generalizations of the model to account for dislocation multiplication - which together with an appropriate choice of the concomitant fit parameters would make the model perform much better in the following benchmark exercise. Here, however, we are interested in the model precisely because of its very simple and transparent form which does not contain any ad hoc assumptions regarding dislocation kinematics.

\subsubsection{The model of Arsenlis: screw-edge representation of dislocation kinematics}
\begin{figure}
	\centering
	\includegraphics[width=.75\linewidth]{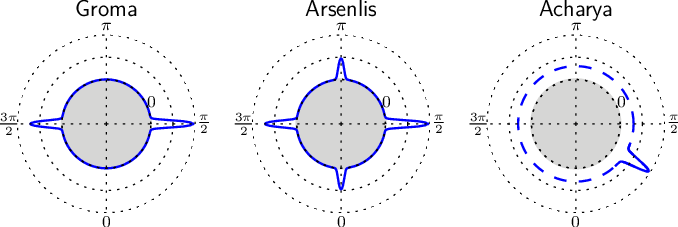}
	\caption{Schematic depiction of the DODF implied by the Groma, Arsenlis and Acharya models. Note that the peaks symbolize Dirac delta functions. Left: Groma $\pm$ uses edge dislocation densities of both signs. center: Arsenlis describes the dislocation microstructure with four densities of $\pm$ edge and screw orientations. right: In Acharya's model the dislocations are divided into a GND density with known direction and SSDs whose orientation distribution is unspecified.}
	\label{fig:DODF_GAA}
\end{figure}

\citet{Arsenlis2004_JMPS52} explicitly acknowledge the distinction between dislocation kinematics and dynamics. In order to make the kinematic problem  computationally more tractable, they restrict the dislocation orientation degrees of freedom to a small number of line directions by taking into account only the four orientations $\rhoep$, $\rhoem$, $\rhosp$, $\rhosm$, where superscripts $e$ and $s$ denote either pure edge or pure screw densities, respectively, and the sign ($+$) or ($-$) indicates the polarity of the dislocation density (\figref{fig:DODF_GAA}). In the 1.5D case and assuming that all dislocations of all orientations have the same mobility and thus the same velocity in a given point, the evolution equations reduce to
\begin{align}
\label{eq:Arsenlis}
\partial_t\rhoep &= -\partial_{x} \left((\rhoep+f^e\rhos)v\right)+ q^s v\nonumber\\
\partial_t\rhoem &= \partial_{x} \left((\rhoem+(1-f^e)\rhos)v\right)+q^s v\nonumber\\
\partial_t\rhos &= (q^{e+} + q^{e-}) v
\end{align}
where $ \rhos= \rhosp+ \rhosm$ and the function $f^e$ (for explicit expression see \citet{Arsenlis2004_JMPS52}) is needed to ensure that dislocation densities do not become negative in the course of evolution. In our presentation of these equations
we have emphasized the analogy with CDD theories by introducing the notations $q^{e+}=
\rhoep/l^{e+}, q^{e-}= \rhoem/l^{e-}$, and $q^{s}= \rhos/l^{s}$ where $l^{e+}$, $l^{e-}$ and $l^{s}$ are average dislocation segment lengths . These quantities play in the Arsenlis formulation very much the same role in controlling dislocation multiplication as the curvature density in CDD theories. Accordingly, 
\citet{Arsenlis2004_JMPS52} complement the equations \eqref{eq:Arsenlis} by equations of evolution for $l^{e+}$, $l^{e-}$ and $l^{s}$. These equations are the counterpart of the equation for curvature density evolution in CDD theories. We do not give them explicitly since in the absence of dislocation sources -- which is the case in the present benchmark problem where the number of loops is fixed -- the equations given by \citet{Arsenlis2004_JMPS52} imply that $q^{e+}, q^{e-}$ and $q^{s}$ are time independent such that we can replace them by their initial values. 

To be consistent with the initial values of the dislocation density and dislocation density vector in our DDD simulations we set the initial dislocation densities as $\rho^{e\pm} = \rhos/2 = \rho_0/4$.  
Furthermore, to correctly reproduce the initial dislocation line length increase we set $q^{e\pm}=q^{s}/2=\rhot/(4r_0)$. For purposes of comparison with the other theories we consider the total dislocation density $\rhot = \rhoep + \rhoem + \rhos$,  the edge component of the GND dislocation density vector, $\AItwo = \rhoep - \rhoem$, and the mobile SSD density $\rho^{\text {SSD}} = \rhot - |\AItwo|$. 

\subsubsection{Models of Acharya and co-workers: Combining the dislocation density tensor with phenomenological models for statistical dislocation density evolution}

The crystal plasticity model of  \citet{Acharya2006_JMPS} starts out from the classical kinematic equation for GND transport formulated by \citet{Mura1963_PhilMag}, $\partial_t\Balpha  = -\nabla \times (\Balpha\times \Bv)$ where $\Bv$ is the velocity vector of the geometrically necessary dislocations. In order to account for the contribution $\BL_p$ of statistically stored dislocations to the plastic distortion rate, this is generalized to 
\begin{align}
\label{eq:acharya}
\partial_t\Balpha  &= -\nabla \times (\Balpha\times v +\BL_p),
\end{align}
where $\BL_{p}$ is the plastic distortion rate due to the motion of SSDs. Acharya and co-workers do not make attempts to derive the temporal evolution of $\BL_p$ from the underlying motion of dislocations, but rather consider this as a problem which should be solved in the spirit of classical continuum mechanics by making constitutive assumptions on a phenomenological basis. 

We consider two cases: (i) the constitutive assumptions made by \citet{Acharya2006_JMPS} (henceforth referred to as 'Acharya') suggest to set
\begin{align}
\label{eq:rho_m}
\BL_p &=  \BP \rho^{\text m} b v.
\end{align} 
where $\BP = (\Bb \otimes \Bn)/b$ is the slip system projection tensor and $\rhom$ is implicitly assumed to account for the statistically stored, mobile dislocations only. Further assumptions in the same paper 
are tantamount to using the same velocity modulus $v$ for geometrically necessary and statistically stored dislocations (very reasonable given that any individual  dislocation has no means of deciding whether it is 'geometrically necessary' or 'statistically stored') and considering the density of statistically stored mobile  dislocations as constant.  \\

Fressengeas, Varadhan and Beaudoin (\citet{Fressengeas2005}, henceforth referred to as FVB) instead proposed to derive the evolution of the mobile SSD density from the phenomenological Kubin-Estrin model (see e.g. \citet{Kubin2002}) which yields:
\begin{align}
\label{eq:Fresengeas}
\partial_t \rhom &= [C_1/b^2 - C_2 \rhom - C_3/b \sqrt{\rho_\text f}] \dot{\gamma} 
\end{align} 
where $C_1$ accounts for mobile dislocation multiplication, $C_2$ accounts for mobile dislocation annihilation and/or dipole formation, and the term with $C_3$ accounts for immobilization by forest obstacles. $\dot{\gamma}$ is the total plastic shear strain rate in the slip system and includes both SSD and GND contributions. Since forest obstacles are absent in our benchmark example which considers dislocations on a single slip system, we only consider the first two terms.  (Note that considering a dipole formation term is not inconsistent with our assumption of a quasi-stationary arrangement of dipole-like obstacles, since in actual PSB microstructures the dipole densities exceed the mobile dislocation densities by about two orders of magnitude and change little over one loading cycle). The parameters $C_1$ and $C_2$ are fitted \textit{bona fide} to reach, at the end of the simulation, the same average plastic strain and mobile dislocation density as found in our benchmark hdCDD theory. 

In both cases (i) and (ii) $\Balpha$ corresponds to $\AI$ via $\Balpha = \AI \otimes \Bb$ and the total density $\rhot$ is equal to $\rhom +|\AI|$. Accordingly we set the initial values $\rhom = \rho_0, \Balpha = 0$. For comparison with the other theories we consider the total dislocation density $\rhot$,  the edge component of the GND dislocation density vector, $\AItwo = \alpha_{21}/b$, and also the mobile SSD density $\rhom$. 

\subsection{Results from DDD}
\label{subsec:benchmark_DDD}

\begin{figure}
	\centering
	\hbox{}\hspace{2cm} $\gamma=0$ \hfill $\gamma=0.33\%$ \hfill $\gamma = 1\%$\hspace{2cm}\hbox{}\\[0.5em]
	\includegraphics[viewport=155 50 480 550, clip,width=0.31\linewidth]{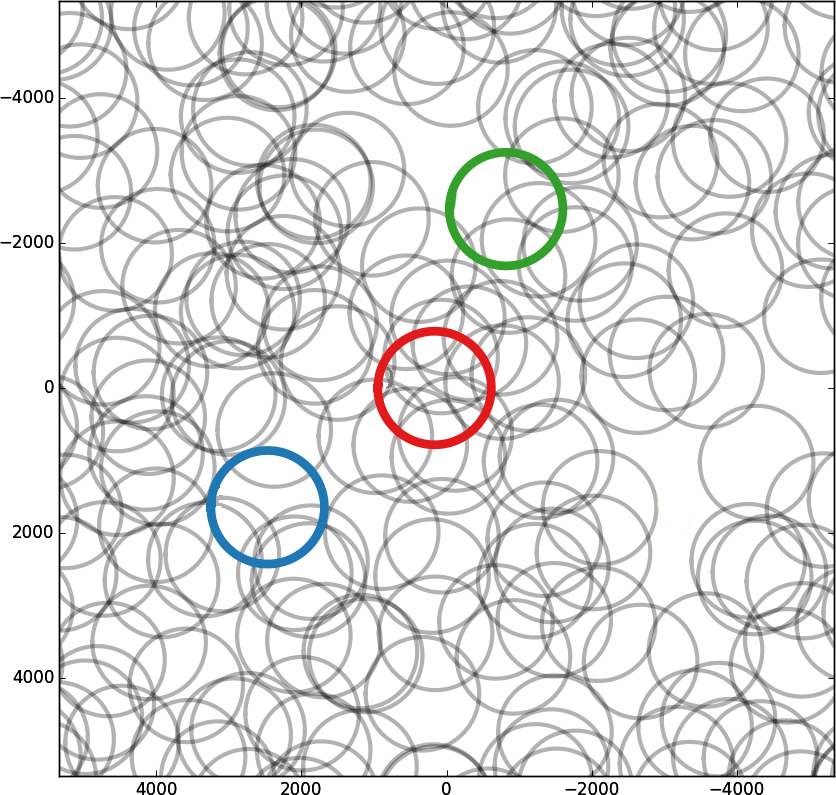}\hfill%
	\includegraphics[viewport=155 50 480 550, clip,width=0.31\linewidth]{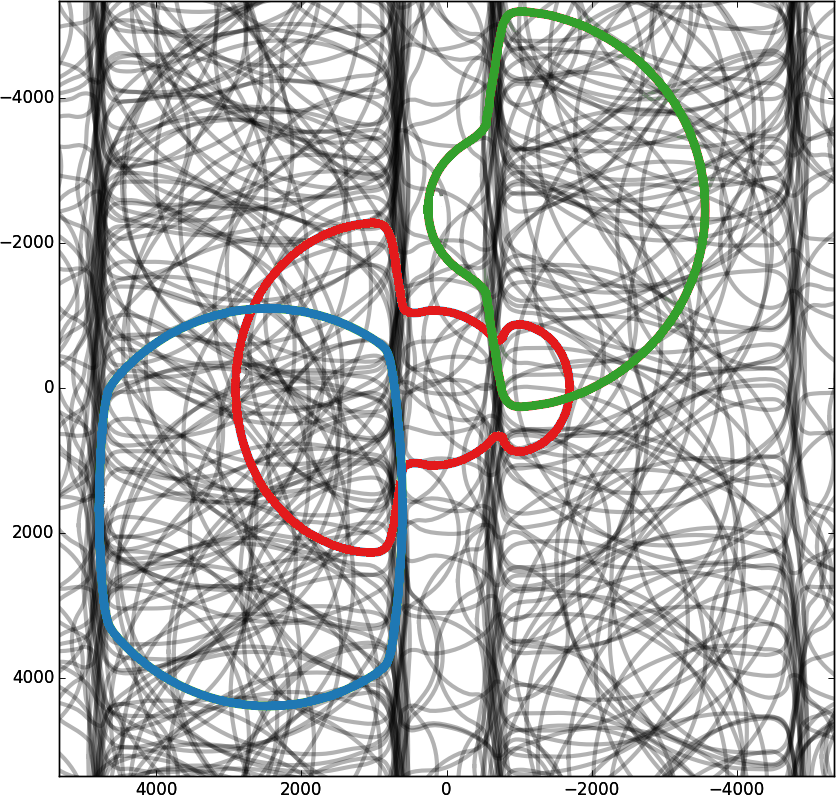}\hfill%
	\includegraphics[viewport=155 50 480 550,
	clip,width=0.31\linewidth]{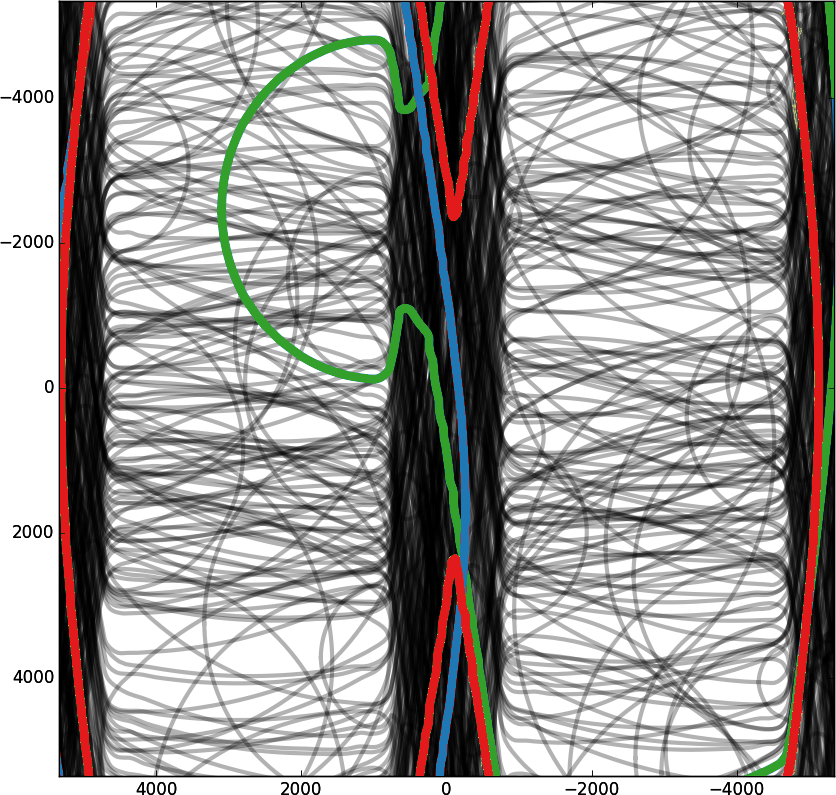}\hfill%
	\caption{Evolution of an initially random distribution of discrete dislocation loops in the 'PSB' velocity field shown in \figref{fig:PSB} . From left to right: initial values and dislocation microstructure at two subsequent time steps, corresponding to strains of $0.33 \%$ and $1\%$. }
	\label{fig:DDD1}
\end{figure}

Reference data are obtained from DDD simulation with an initially random distribution of loops. \Figref{fig:DDD1} shows typical dislocation microstructures at three different time steps. Note that the loop configuration is projected along the $z$ axis into the $xy$-plane: The different loops expand on different slip planes.
A number of important features can be observed: dislocation loops in the channel expand until they reach the wall. There, line segments change their orientation such that they align with the contour lines of the velocity field. Parts of dislocation loops in the channel regions further expand until they reach near-screw orientation and become almost straight. Since the wall regions are not completely impenetrable, dislocations eventually cross the wall and leave on the other side to bow out in the adjacent channel. A significant number of dislocations is getting 'trapped' inside the walls and form near-parallel edge dislocation bundles of zero net Burgers vector. In the transition region between the wall and the channel we observe that the radius of curvature, with which lines are bent, is smaller than the initial loop radius. 

\subsection{Results from hdCDD and {CDD}$^{(1)/(2)}$ vs. DDD: analysis of dislocation density in the configuration space}
\label{subsec:benchmark_CDD}

\begin{figure}[htp]
	\centering
	\includegraphics[width=0.90\textwidth]{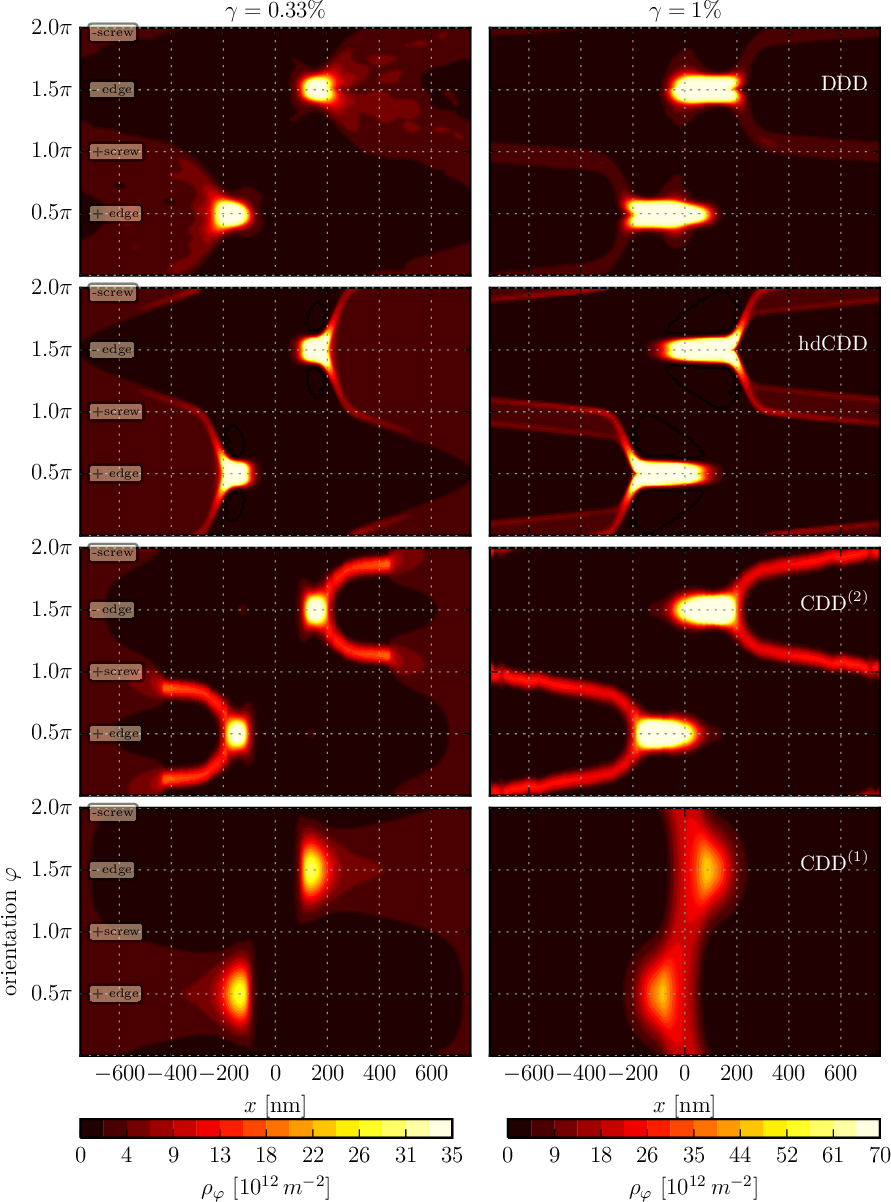} 

	\caption{\label{fig:hdCDD}
		hdCDD density $\rho(x,\vp)$ in the configuration space: 
		the vertical axis denotes the line orientation, the horizontal axis is the x-axis; top row: $\rho(x,\vp)$ as reconstructed from DDD,
		second row: $\rho(x,\vp)$ as obtained by solution of the hdCDD equations, third and fourth rows: $\rho(x,\vp)$ as reconstructed from the 
		solutions of the \CDDI and \CDDII equations.}
\end{figure}

For a detailed comparison of the DDD microstructure with CDD results we convert the discrete lines into a continuous density in the higher-dimensional hdCDD configuration space, here spanned by the variables $x$ and $\varphi$. This offers the possibility to directly compare with hdCDD, and with the reconstructed angular densities derived from \CDDI\ and \CDDII. Since the benchmark system is statistically homogeneous in $y$ and $z$ directions, the angular densities depend on $x$ only. We obtain continuous data by first applying the \emph{D2C} conversion to the three-dimensional DDD data, followed by averaging over the $y$ and $z$ directions. Higher-dimensional density distributions which result from this process at strains of $0.33\%$ and $1\%$ are shown in the top row of \figref{fig:hdCDD}. 

In principle it is possible to use the spatially averaged initial DDD microstructure directly as initial condition for the continuum simulations \citep[compare][]{Sandfeld2015b_MSMSE}. Instead, we consider as the initial state the statistically homogeneous and isotropic densities which correspond to an average over the initial conditions in an infinite ensemble of DDD simulations, or to the spatial average obtained for a system of infinite extension in $y$ and $z$ directions. This leads to the following initial conditions: The hdCDD dislocation density and curvature density functions are isotropic and spatially homogeneous, $\rho\rphi = \rho_0/(2\pi), q\rphi = \rho_0/(2\pi r_0)$. The corresponding initial values for the \CDDI\ and \CDDII\ variables are $\rhot = \rho_0,\qt = \rho_0/r_0, \AI = 0, \AII = \rhot/2 \BI^{(2)}$. For hdCDD, the evolved density function $\rho\rphi$ can be directly compared to the DDD data. For \CDDI\ and \CDDII\ we evaluate the DODF from the field variables in each spatial point using the maximum entropy principle as discussed in \secref{subsec:CDD2} and then multiply with $\rhot$ to obtain the matching density function $\rho\rphi$.

At first dislocation loops expand uniformly in the channels ($x\approx \pm(300\ldots 750)$ nm), where in the hdCDD configuration space positive edge orientations ($\varphi=0.5\pi$) move towards the right and negative edge orientations ($\varphi=1.5\pi$) move towards the left. Eventually more and more density from the channel reaches the wall where positive and negative edge components of dislocations accumulate on the left and right sides of the low velocity region, respectively. These accumulations show in \figref{fig:hdCDD} as spots of high hdCDD density. We observe that both hdCDD and \CDDII\ are in very good agreement with the reference DDD simulations in terms of the morphology and location of the density accumulations. \CDDI\ also shows accumulation of GND in edge orientation at the walls but fails to capture the concomitant transition to screw orientations in the channels. 

At a strain of $0.0033$  the channels are already partially depleted of edge dislocations which accumulate at the left and the right edge of the wall whereas the dislocation density at the center of the wall (around $x=0$) is still low because the small velocity makes it difficult for dislocations to penetrate the wall. This is well captured by all models. Dislocation loop segments approaching the wall assume more and more straight edge-like orientations, whereas segments moving in the channels are preferentially screw oriented. hdCDD is able to represent this feature exactly, and \CDDII\ shows qualitative agreement, while \CDDI\ does not indicate any enhancement of near-screw orientations in the channels.  

When the system further evolves (strain 0.01), edges align inside the wall into dipole bundles. hdCDD correctly represents this through the two high density accumulations which now overlap in the spatial projection. In accordance with \figref{fig:PSB}(a) and the DDD plot in \figref{fig:DDD1} the curvature of these dislocations is almost zero. At the same time in the channel the density consists almost exclusively of threading dislocations in positive and negative near-screw orientations. Both the DDD data and the hdCDD simulations show this feature which is also correctly represented in the \CDDII\ simulations. \CDDI\, by contrast, makes the erroneous prediction that the channels are depleted of edge {\text and} screw dislocations.

We conclude that \CDDI\ is not able to fully predict the microstructure evolution. The segregation of edge dislocations in the walls and screw dislocations in the channels is not captured. \CDDII\, on the other hand, is able to represent most important characteristics of the PSB-like system. Only small details show deviations from DDD. Finally, hdCDD is able to reproduce all relevant mechanisms and the time evolution of dislocation microstructure in very good agreement with DDD. This confirms the observation made in previous works that hdCDD is kinematically exact, i.e., its results coincide with averages over an infinite ensemble of matching DDD simulations carried out assuming the same velocity field. 

DDD, \CDDII and hdCDD predictions are in a good qualitative agreement with observations of dislocation microstructure and dislocation motion in PSB in copper using transmission electron microscopy (TEM) (\figref{fig:PSB_Mughrabi}). We quote \citet{Mughrabi1979}  who stated that ``(1) edge dislocations bow out of the walls and become entangled in the neighboring walls; (2) edge dislocations that are newly formed in the wake of moving screw dislocations are incorporated into the wall''. We note that \citet{Mughrabi1987_pssol} also reports the observation that dislocations with the same orientation, located at equivalent positions in the wall-channel structure of a PSB, have the same curvature -- which is a main assumption for curvature-density fields in the hdCDD and CDD theories. Also the observation that this curvature is high near the wall edges but lower in the channels is in line with the results of our simulations, see \figref{fig:hdCDD}.

\subsection{Comparison of results from {CDD}$^{(2)}$ with other theories: dislocation density/strain profiles and average behaviour}

\begin{figure}[hp!]
	\centering
	\includegraphics[width=0.95\textwidth]{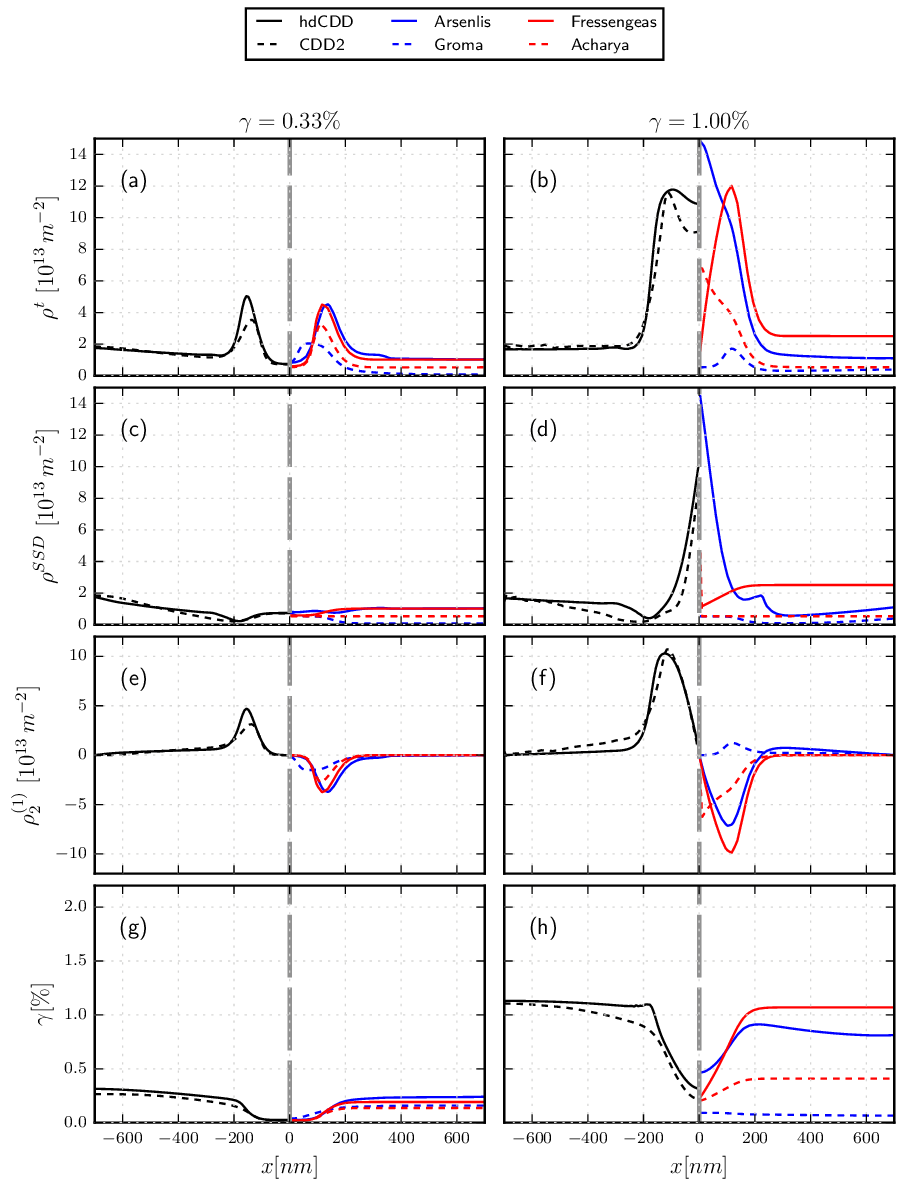}
	\caption{Evolution of total dislocation density $\rhot$, mobile SSD density $\rho^{\text {SSD}}$, edge GND density $\AItwo$, and plastic slip $\gamma$ in the slip plane.  Using the (anti)symmetry of the system hdCDD and \CDDII are plotted only on the left side of the domain  and the other models on the right side.}
	\label{fig:PSB_line}
\end{figure}

To compare \CDDII with other theories we need to specify a minimum set of variables which can be found in all of the theories considered. In what follows we look at the spatio-temporal evolution of the total dislocation density $\rhot$, the edge component of the GND vector $\AItwo$ and  the plastic slip $\gamma$. We also include a derivative quantity, namely the mobile SSD density which can be evaluated as $\rho^{\text {SSD}} = \rhot - |\AItwo|$. These quantities are shown in \figref{fig:PSB_line} for the same two snapshots in time as in the previous section. For hdCDD the integrated densities $\rhot$, $\AItwo$ are obtained from the higher-dimensional $\rho\rphi$  using \eqref{eq:rhot}-\eqref{eq:AII}, \CDDII\ already contains these field variables in explicit form, whereas from the other theories $\rhot$ and $\AItwo$ can be deduced as discussed in the respective sections. 

We observe in \figref{fig:PSB_line}(e) that the initial increase in GND density at the wall boundaries is predicted by all models. Models without dislocation multiplication (Groma) or which associate dislocation density increase only with GNDs (Acharya), however, predict a slightly lower GND density concomitant with a lower strain. Comparing the edge GND density $\AItwo$  with the total density  (\figref{fig:PSB_line}(a)) one observes that at the boundaries of the wall all dislocations have edge character which is also predicted by all models. Most of the plastic strain is generated inside the channel  where the average dislocation velocity is higher. 

However, as the systems evolve further, only \CDDII\ and the Arsenlis model show good qualitative agreement with the hdCDD or DDD results (\figref{fig:PSB_line}(b)), while models without dislocation multiplication (Groma, Acharya) strongly underestimate the strain in the channel and the dislocation density in the wall. An intermediate position is provided by the FVB model with dislocation multiplication. This reproduces accurately the strain and GND density profiles but strongly underestimates the density of SSDs in the wall -- where it predicts a minimum rather than a maximum of the SSD density. This feature is important because it makes the model a poor candidate for explaining wall formation - which requires SSDs to form narrow dipole configurations inside the walls rather than in the channels. According to the FVB model with dislocation multiplication, the opposite is true since the mobile SSD density (the density of those dislocations which can mutually trap into dipoles) is highest in the center of the channel and lowest in the wall. 

We may also look at averages. All models which account for dislocation multiplication can more or less correctly reproduce the increase in average plastic strain, and the concomitant increase in the spatially averaged total dislocation density. In the case of \CDDII\ and the Arsenlis model this is achieved without parameter fitting, in case of the FVB model we have fitted the multiplication and dipole formation parameters to obtain the correct values of end strain and mobile dislocation density. 
\begin{figure}[h!]
	\centering
	\includegraphics[width=\textwidth]{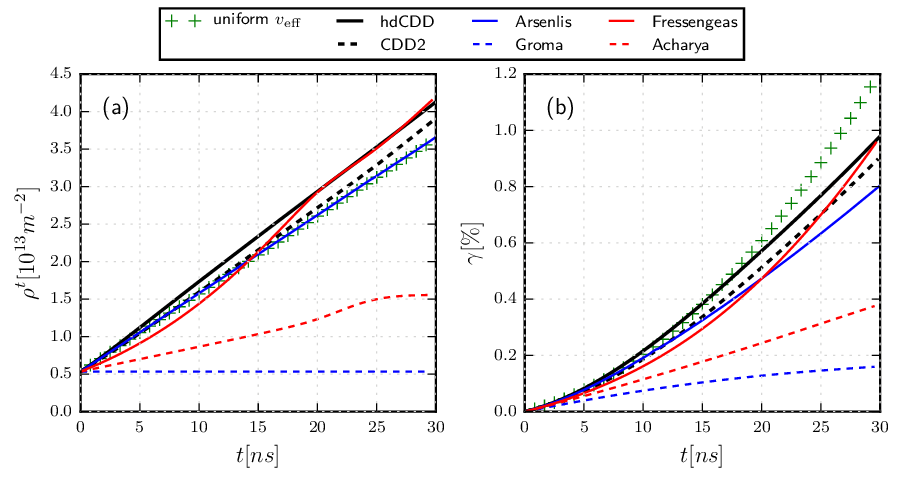}	\caption{\label{fig:PSB_time}Normalized total dislocation density \rhot and total plastic slip $\gamma$ over time.}
\end{figure}
We  observe two distinct stages in the evolution of $\rhot$ and $\gamma$: in the first stage ($\gamma < 0.33\%$) we find free expansion of loops which leads to a linear increase of the average dislocation density  $\langle\rhot\rangle$ (which is proportional to the circumference of all loops) and a quadratic increase of the average plastic strain $\langle\gamma\rangle$ (which is proportional to the area of all loops). In this stage most theories can predict the evolution of dislocation density more or less correctly. In a second stage deformation is mainly by screw dislocations moving in the channels while dislocation multiplication is mainly by the concomitant drawing out of edges along the wall. In this stage, both the strain and the dislocation density increase with time in an approximately linear manner.  

An instructive feature shows up when we compare the amount of dislocation multiplication with that in a spatially homogeneous reference system where dislocations move at an effective velocity $v_{\text {eff}}=\langle v(x)\rangle$ which is the spatial average of our homogeneous velocity field. Surprisingly, both hdCDD and to a lesser extent \CDDII predict that the heterogeneous system shows a larger dislocation density increase than the homogeneous reference system with uniform velocity (\figref{fig:PSB_time}(a)), whereas the total plastic slip generated by the movement of dislocations is less than for the homogeneous reference system (\figref{fig:PSB_time}.b). This is due to the following features: (i) dislocations accumulate in the wall where they move more slowly than on average, leading to a reduced total strain, (ii) dislocation segments which have crossed a wall can expand in the adjacent channel and effectively act as additional sources, leading to enhanced multiplication. The Arsenlis model which does not account for spatial transport of dislocation curvature predicts the dislocation increase in the heterogeneous system to be identical to the uniform reference system. This indicates that it can not fully account for the effects of  dislocations bowing out from a wall. The FVB model produces a correct description of the strain and dislocation density increase - two variables to which the two parameters of the model have been fitted. The Groma and Acharya models, which cannot account for dislocation multiplication where this leads to accumulation of statistically stored dislocations, by contrast, completely underestimate \rhot and $\gamma$.

\section{Conclusion}

We have applied the Maximum Information Entropy Principle (MIEP) to reconstruct the Dislocation Orientation Distribution Function (DODF) from a finite set of the associated alignment tensors and have used this reconstruction for kinematic closure of CDD evolution equations. This method can in principle be used for closure of CDD theories using alignment tensors of any order, though we have explicitly considered only the lowest two orders. 

For testing the performance of the resulting evolution equations, we have proposed a generic approach which compares averaged DODF obtained from DDD simulations with those reconstructed from CDD models. We have carried out such a comparison for a benchmark example, namely the motion of dislocations in an idealized PSB microstructure. The comparison demonstrated the capabilities of boh hdCDD and \CDDII\ to account for all essential features of dislocation motion in such microstructures. At the same time \CDDI\ -- which performs well in other situations, see e.g. \citet{Hochrainer2014JMPS} -- was shown to be incapable of dealing with situations where edge-screw asymmetry is pronounced. This demonstrates that the formulation of CDD evolution equations requires careful consideration of the structural complexity of the problem under consideration. 

We have then benchmarked the performance of \CDDII\ against several models published in the literature. 
Out of the investigated models only \CDDII\ and the Arsenlis' model -- which is specifically designed to capture screw-edge systems -- were able to predict all essential characteristics of dislocation motion in the strongly anisotropic PBS microstructure.  

Because of its kinematic exactness, hdCDD -- even though computationally inefficient as a simulation tool -- provides a useful reference for assessing kinematic consistency of simpler theories. Obtaining a similar reference for assessing dynamic consistency is a more difficult task: In this case, we need to perform ensemble simulations of discrete dislocation systems and then use ensemble averaging to determine the corresponding averaged density measures and their time evolution. Performing an assessment of dynamic consistency of the proposed approaches, for instance for interacting loops under an external shear stress in the geometry outlined above, as well as for other benchmark problems, remains an essential task for future investigations. 

\section*{Acknowledgements}
Financial support from the Deutsche Forschungsgemeinschaft
(DFG) through Research Unit FOR1650 'Dislocation-based Plasticity' (DFG grants No~SA2292/1-1 and ZA171/7-1) as well as the European m-era.net project 'FASS' (DFG grant No~SA2292/2) is gratefully acknowledged.


\section*{References}
\bibliographystyle{elsarticle-harv}
\bibliography{./CDDMaxEnt}

\newpage
\appendix

\section{Comparison of computational cost of different density-based theories}

One of the possible advantages of coarse-grained continuum dislocation theories over discrete dislocation dynamics is their lower computational cost. The number  of degrees of freedom in a continuum dislocation model equals the number of state variables times the number of nodal points used in the representation of the density fields.  Hence the number of state variables  in each spatial point is a good criterion for comparing the computational cost of different continuum models.  In our 1.5D benchmark  test,  the homogeneity of the problem in the direction of screw dislocation glide implies that the geometrically necessary dislocations must always have edge orientation which implies that $\AIone=0$, $\AIIonetwo=0$ and $\rhosp=\rhosm$ which reduces the number of state variables. In a fully 3D formulation,  \AI  and \AII remain 2D vectors and symmetric tensors in the local slip system coordinates and therefore have 2 and 3 components, respectively. In the higher-dimensional hdCDD theory, the variables $\rho_\vp,q_\vp$ have to be partitioned in the orientation space in addition to the spatial domain  which makes hdCDD infeasible for real 3D systems. In the present benchmark problem, we use for hdCDD a partition of orientation space into $n_\vp=60$ regular Galerkin elements. Comparing the total number of  DOF and state variables for each model given in Table (\ref{tab:models}) shows that all the models except hdCDD have relatively low computational cost.
\begin{table}[h]
	\centering
	\begin{tabular}{c|cclcccl}
		&          hdCDD           &        \CDDII         & \CDDI                  &             Arsenlis             &        Groma         &   Acharya    & FVM                \\ \hline
		variables & $\rho_\vp,q_\vp,\gamma $ & $\AI,\AII,\qt,\gamma$ & $\AI,\rhot,\qt,\gamma$ & $\rho^{e\pm},\rho^{s\pm},\gamma$ & $\rho^{e\pm},\gamma$ & $\AI,\gamma$ & $\AI,\rhom,\gamma$ \\
		1.5D/2.5D &        $2n_\vp+1$        &           5           & 4                      &                4                 &          3           &      2       & 3                  \\
		3D     &        $2n_\vp+1$        &           7           & 5                      &                5                 &          -           &      3       & 4                  \\
		DOF   &          $\sim48000$          &         $\sim 2000$    & $\sim 1600$        &     $\sim 1600$              &       $\sim   1200$         &     $\sim 800$     & $\sim 1200$
	\end{tabular} 
	\caption{		\label{tab:models}
		Number of state variables per slip system at each spatial point and total degrees of freedom in the benchmark problem for different models. The domain is discretized with 200 second order elements.}
\end{table} \\

If we now compare with our DDD reference model we may calculate the number of DOF for the reference model as follows: In the reference model we have 900 loops, and in order to achieve an angular resolution comparable to hdCDD each loop needs to be discretized into 60 segments (corresponding to an initial nodal spacing of about 10nm). This corresponds to a total of 54000 nodal DOF. However, we emphasize that this is not a fair comparison since the continuum calculations use the statistical homogeneity of the benchmark problem in $y$ and $z$ directions to reduce  dimensionality which obviously results in a tremendous reduction of degrees of freedom -- for a generic three-dimensional problem 
of the same size the number of DOF would need to be about three orders of magnitude higher. On the other hand, we also emphasize that the main differences in computational cost between DDD and coarse-grained continuum theories arise from the possibility, in a density-based theory, to represent segment-segment interaction stresses as local functionals of the dislocation densities, instead of calculating them by summation over all segment pairs, whereas the calculation of long range stresses can be carried out on a much coarser mesh than is needed for solving the transport equations. We will discuss these issues in a future work which presents a three-dimensional dynamical implementation that includes interaction stresses. 

\section{Dislocation annihilation in continuum dislocation dynamics}

The problem of annihilation of dislocations only emerges in coarse-grained continuum theories that allow for the coexistence of dislocations with different orientation within the same volume element. CDT theories such as \cite{Arsenlis2004_JMPS52}  formulate this problem in a conceptual framework that focuses on encounters of straight lines which annihilate once they meet within a reaction cross-section (annihilation distance) leading to bi-molecular annihilation terms. But dislocations are not particles. Dislocations annihilate when two loops merge - a process which in itself consumes very little dislocation line length - and the subsequent motion then may lead to a reduction rather than an increase in line length, but at other locations and in different orientations from those of the annihilating segments.  

Experimental observation indicates that the annihilation of segments with near-screw orientation may proceed via a cross slip mechanism and that the minimum distance $y_{\rm a}$ required for this process to occur is significantly (by about two orders of magnitude) larger than for annihilation of segments of other orientations. Therefore we first consider situations where the annihilation process is initiated by cross slip and subsequent annihilation of two near-screw segments which are oriented at a small angle $\vp_{\rm a} \in [- \Delta \vp, \Delta \vp]$ from the screw orientations $\vp_{\rm a} = 0$ and $\vp_{\rm a}=\pi$.

\begin{figure}[t]
	\centering
	\includegraphics[width=0.95\linewidth]{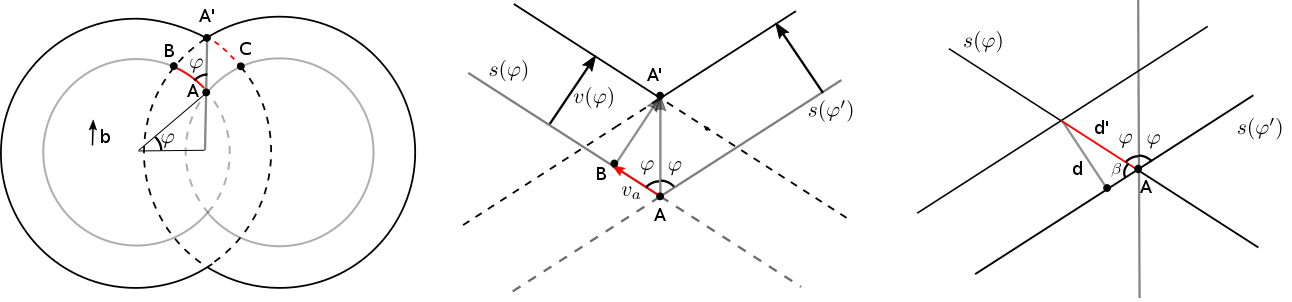}
	\caption{Top view of a cross slip annihilation process. Left: Cross slip initiates the  annihilation of two merging dislocation loops.  Gray and black lines depict the evolution of the merging loops during a small time step. The dashed lines shows the annihilated parts of the loops.  Center: The intersection point $A$ moves to a new position $A'$. Its relative motion to the segment $\vp$ decreases the segment length by the length $AB$. Right: segments with orientation $\vp'$ have expected spacing $d$ perpendicular to their line direction. If evaluated along the line direction $\Bl(\vp)$, their spacing is $d'=\frac{d}{\sin(\beta)}$.}
	\label{fig:annihilation_loop}
\end{figure}

\figref{fig:annihilation_loop} (left) depicts a situation some time after near-screw segments of two loops moving on parallel slip planes with normal vector $\Bn$ have merged by cross slip.  As the loops expand, the intersection point $A$ -- which corresponds to a segment in the cross slip plane and separates segments of direction $\Bl(\vp)$ and $\Bl(\vp')$ in the original slip planes -- moves in the Burgers vector direction to the new position $A'$. During this movement of the intersection point the segment $AB$ with orientation $\vp$ as well as a matching segment with orientation $\vp' = \pi - \vp$ (\figref{fig:annihilation_loop}, center) rotate into the respective screw orientations and annihilate. Assuming that the loops have the same curvature and that the dislocation velocity is independent of the segment orientation $\vp$, the annihilation process is symmetric with respect to the direction of the Burgers vector. 

The total annihilation rate is determined by the probability of finding, for a given segment of orientation $\vp$, an intersection point with a matching segment of orientation $\pi - \vp$, and by the velocity of this intersection point parallel to the segment direction $\Bl(\vp)$. As depicted in \figref{fig:annihilation_loop} (centre), if the segments of both loops are moving with velocity $v$, then the velocity component of the intersection point parallel to $\Bl(\vp)$ is
\begin{align}
v_{\rm a}&=v\cot(\vp)
\end{align}
To evaluate the probability of finding for a segment of orientation $\vp$ an intersection point with a matching segment we proceed as follows: (i) Matching segments must have orientations between $\pi - \vp - \Delta \vp$ and $\pi - \vp + \Delta\vp$. The area density of intersection points of such segments with a perpendicular plane (normal vector $\Bl(\pi - \vp)$) is given by 
\begin{align}
\rho_{\rm a}(\pi - \vp) = \int_{\pi - \vp - \Delta \vp}^{\pi - \vp + \Delta\vp} \rho(\vp') {\rm d}\vp'.
\end{align} 
(ii) We now consider a strip of width $2 y_{\rm a}$ around the slip plane of the first segment, but still contained in the plane with normal vector $\Bl(\pi - \vp)$. The average distance $d'$ between intersection points, evaluated along the direction of this strip, follows from the condition $2 d' y_{\rm a} \rho_{\rm a} (\pi - \vp) = 1$. Their spacing along the direction $\Bl(\vp)$ is then (see \figref{fig:annihilation_loop}, right)
\begin{align}
d(\pi - \vp,\vp)= \frac{d'}{\sin(\beta)}=\frac{d'}{\sin(2\vp)}\quad,\quad d'=\frac{1}{2 \rho_{\rm a} y_{\rm a}}. 
\end{align}
We now assume that the configurations of the two merging loops are statistically independent. This assumption implies that the loop centers are located at statistically independent points and that each loop expands, away from the intersection point, in a manner that is not strongly influenced by the presence of the second loop. In that case the volume density of such intersection points can be evaluated as the line length of orientation $\vp$ per unit volume, divided by the mean spacing $d(\vp,\pi - \vp)$ of intersection points along a segment of this orientation. Accordingly, we obtain
\begin{align}
\rho_\text{int}(\vp,\pi -\vp) =\frac{\rho(\vp)}{d'}= 2 y_{\rm a} \sin (2\vp)
\left[\rho(\vp)  \int_{\pi - \vp - \Delta \phi}^{\pi - \vp + \Delta\phi} \rho(\vp') {\rm d}\vp' \right] 
\end{align}
It follows that the temporal change of $\rho(\vp)$ due to the motion of such intersection points ('annihilation rate') becomes:
\begin{align}
\left.\partial_t \rho(\vp)\right|_{\rm anil} &= - 2 \left[\rho(\vp) \int_{\pi - \vp - \Delta \phi}^{\pi - \vp + \Delta\phi} \rho(\vp') {\rm d}\vp' \right] \frac{\cos(\vp) }{\sin(\vp)} \sin (2\vp) y_{\rm a} v \\
&= -4 \left[\rho(\vp)  \int_{\pi - \vp - \Delta \phi}^{\pi - \vp + \Delta\phi} \rho(\vp') {\rm d}\vp' \right] \cos^2 (\vp)y_{\rm a} v  .
\end{align}

\begin{figure}[t]
	\centering
	\includegraphics[width=0.45\linewidth]{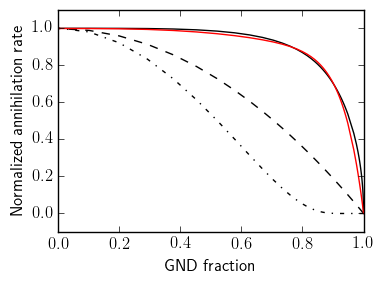}
	\caption{Normalized annihilation rate as a function of GND fraction for screw-triggered dislocation annihilation: full line: annihilation rate for a DODF which is	symmetrical around edge orientation (edge oriented GND), red line: polynomial approximation ($1-.2x^3-.8x^{16}$) to full black line, dash-dotted line: result for a DODF which is symmetric around screw orientation (screw oriented GND), dashed line: Parabolic dependency representing the kinetic theory result for parallel screw dislocations.}
	\label{fig:annMI}
\end{figure}

We consider two limiting cases: First, if all dislocations are screw oriented then $\rho(\vp) = \rho_+ \delta(\vp) + \rho_- \delta(\pi - \vp)$. The annihilation rate follows as 
\begin{align}
\left.\partial_t \rho_+\right|_{\rm anil} = \left.\partial_t \rho_-\right|_{\rm anil} = 
- 4 \rho_+ \rho_- y_{\rm a} v 
\end{align}
which is the result expected by kinetic theory for 2D particles moving with velocity $v$ in opposite directions and annihilating if they pass within a reaction cross-section $2 y_{\rm a}$. The second limiting case, which is the one relevant to our benchmark example, is that the dislocations have a smooth angular distribution which can be approximated as constant over the small angle interval $2\Delta \vp$. Then, 
\begin{align}
\left.\partial_t \rho(\vp)\right|_{\rm anil} &=  -8 \Delta \vp \rho(\vp) \rho(\pi - \vp) \cos^2 (\vp) y_{\rm a} v .
\end{align}
We see that the screw-triggered annihilation process removes segments of {\em all} orientations with a rate that is largest for screw orientations and goes to zero for the pure edge orientations $\vp = \pi/2$ and $\vp = 3\pi/2$. The concomitant temporal change of the zeroth-order alignment tensor (total dislocation density) can be calculated by integration:
\begin{align}
\left.\partial_t \rhot \right|_{\rm anil} &=\oint \left.\partial_t \rho(\vp)\right|_{\rm anil}  \,\text{d}  \vp 
\end{align}
The DODF of PSB structures is symmetric around the edge orientations $\vp=\frac{\pi}{2}$ and $\vp=\frac{3\pi}{2}$. As a consequence, 
$\rho(\vp)=\rho(\pi-\vp)$ . Using this symmetry property and the DODF given by Eq. \eqref{eq:CDDI_DODF}, the rate of reduction in total dislocation density in \CDDI can be evaluated as   
\begin{align}
\label{eq:AnnEdgeGND}
\left.\partial_t \rhot \right|_{\rm anil} &= (\rhot)^2 \frac{8\Delta\vp}{Z^2} \left[\oint \exp(-2\lambda_1 \sin(\vp))\cos^2(\vp)\text{d}\vp
\right] y_{\rm a}v
\end{align}
where $Z$ is the partition function of the distribution. The first order alignment tensor is not directly affected by the annihilation
process. For completeness, we also consider the case (not relevant to our benchmark example) where the DODF is symmetric around the screw orientations $\vp =0$ and $\vp = \pi$ such that $\rho(\pi - \vp)=\rho(\pi\vp)$. In that case an analogous calculation gives for \CDDI
\begin{align}
\label{eq:AnnScrewGND}
\left.\partial_t \rhot \right|_{\rm anil} &=(\rhot)^2\oint \frac{8\Delta\vp}{Z^2}  \left[\oint \cos^2(\vp)\text{d}\vp \right] y_{\rm a} v
\end{align}
For a completely isotropic dislocation arrangement, $\lambda_1 = 0$ and $Z = 2\pi$, and we obtain in both cases  
\begin{align}
\label{eq:AnnIsoRho}
\left.\partial_t \rhot \right|_{\rm anil}^0 &=(\rhot)^2 \frac{2\Delta\vp  }{\pi} y_{\rm a}v.
\end{align}
In the presence of excess dislocations of one sign, the annihilation rate is a function of the first moment function $\MI = |\rho^{(1)}|/\rhot$
which can be understood as the GND fraction of the total dislocation density. This dependency is illustrated in Figure \figref{fig:annMI} which shows the annihilation rate, normalized by the value at $\MI = 0$, as a function of the GND fraction $\MI$.  

We can see that the annihilation rate decreases monotonically with increasing GND fraction and goes to zero if all dislocations are GND. However, this decrease is not well described by the parabolic dependency expected according to kinetic theory for a system of straight 
parallel dislocations: In edge-dominated microstructures such as PSBs, where GNDs are of edge orientation but annihilation is triggered by
cross slip of screw dislocations, the kinetic theory expression over-estimates the impact of GNDs on the annihilation process, whereas in microstructures with screw oriented GNDs, the impact of GNDs is under-estimated. 

Assuming an equi-convex microstructure, the annihilation rate of the total curvature density can be straightforwardly evaluated from the 
dislocation density annihilation rate:
\begin{align}
\label{eq:Ann_qt}
\left.\partial_t \qt \right|_{\rm anil}  &=  \left.\partial_t \rhot \right|_{\rm anil} \frac{\qt}{\rhot}. 
\end{align}    
The concomitant reduction in dislocation curvature density decreases the elongation (source) term $v\qt$ in the evolution equation of the total dislocation density \eqref{eq:drhotdt_CDD1} -- an effect which has important long-term impacts on the evolution of dislocation  microstructure and may outweigh the direct effect of annihilation. 

\begin{figure}[tb]
	\centering
	\includegraphics[width=0.8\linewidth]{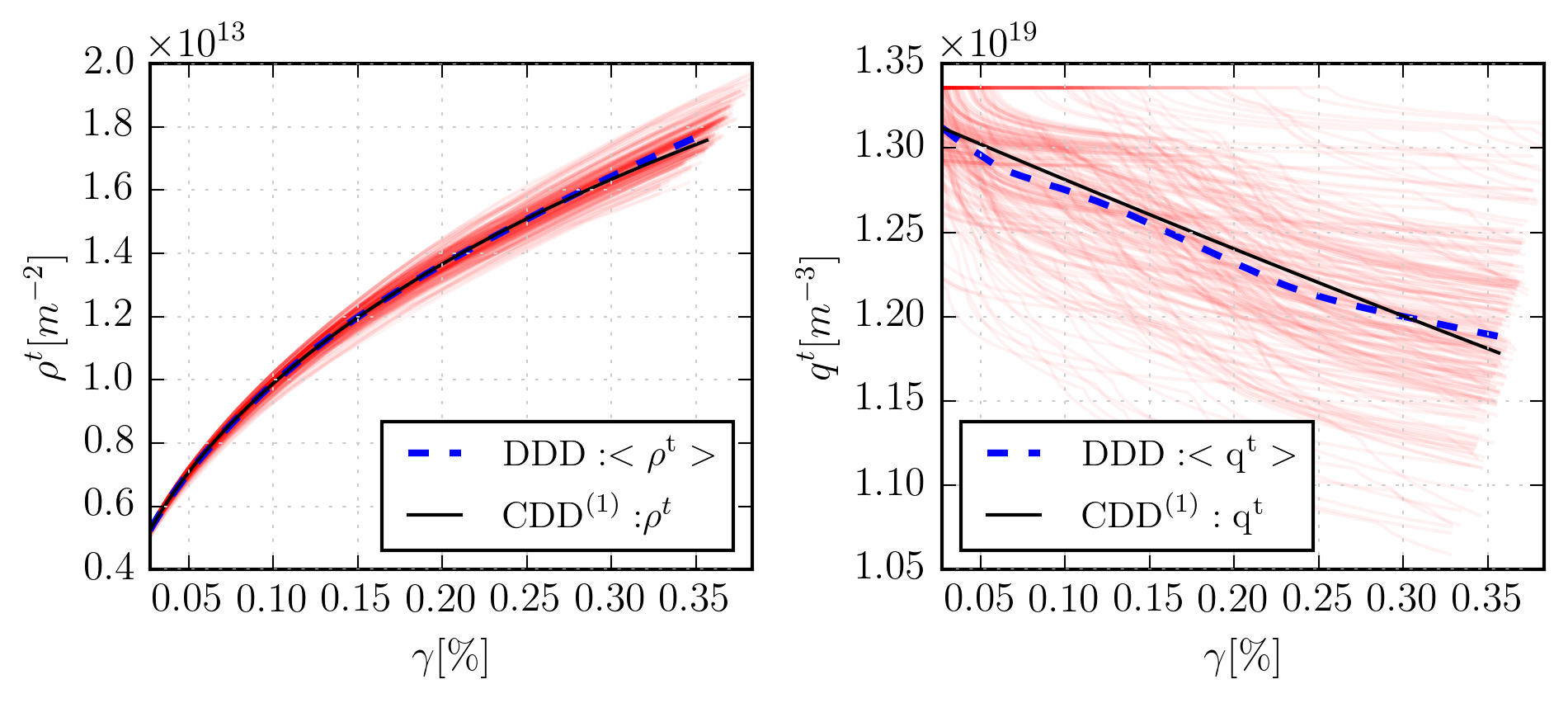}
	\caption{Evolution of total dislocation density and curvature density in a system of loops which annihilate by merging of near-screw
	segments. Dashed lines: averages over 200 discrete simulations, Full lines: CDD predictions, thin lines: evolution in individual DDD runs: for 
	parameters see text.}
	\label{fig:AnnValidate}
\end{figure}

We have validated our model by analyzing the evolution of a system of loops with the same initial radius and dislocation density as in our DDD reference  model. Loops are initially randomly distributed in a periodically continued cuboidal volume and expand with constant velocity $v$. Two loops merge if two conditions are fulfilled: (i) the contacting segments are within an angle $\pm \Delta \vp$ from the screw orientations $\vp = 0,\vp = \pi$. (ii) the spacing of the slip planes on which the loops are positioned is less than $y_{\rm a}$. For the annihilation distance we use the value $y_{\rm a} = 50$ nm given by \cite{Mughrabi1979} as typical for screw dislocation annihilation in PSB microstructures. For the corresponding orientation interval we use the value $\Delta \phi = \pm 15^\circ$ given by \cite{hussein2015}.
The results of the validation exercise shown in \figref{fig:AnnValidate} show an excellent agreement between the average evolution of dislocation density and curvature density as averaged over an ensemble of multiple CDD simulations and the respective predictions based upon Eqs. \eqref{eq:AnnIsoRho} and \eqref{eq:Ann_qt}. 

\begin{figure}[tb]
	\centering
	\includegraphics[width=0.8\linewidth]{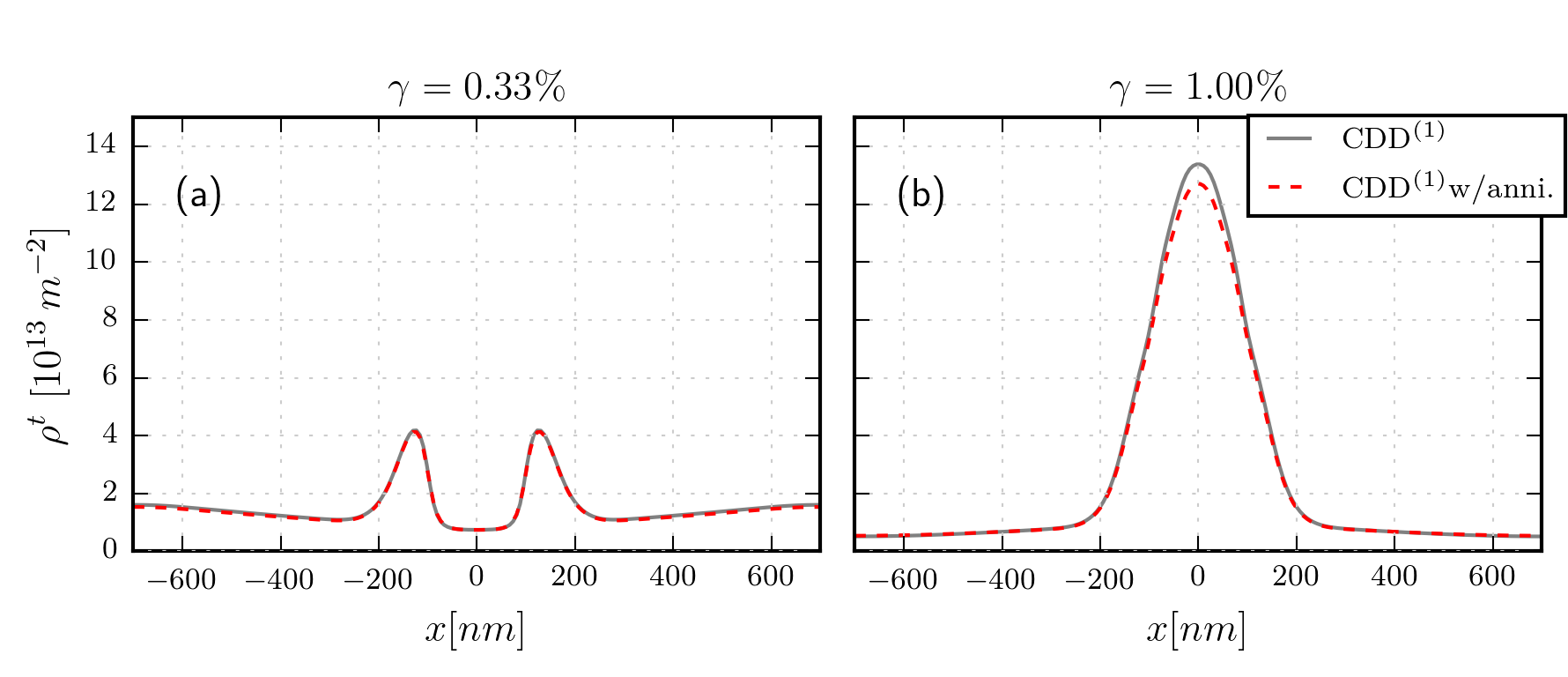}
	\caption{Influence of annihilation  on the evolution of GND density and total dislocation density in CDD1; black line: result without 
	annihilation, red line: result with annihilation. Parameters: $\Delta \phi = \pi/12$, $y_{\rm a} = 50$ nm.}
	\label{fig:CDD1_annihilation}
\end{figure}

To illustrate the influence of annihilation in our benchmark example we have repeated the simulation of the \CDDI model by including the annihilation terms derived above. Parameters used were the same as in the validation exercise. The results shown in figure \figref{fig:CDD1_annihilation} demonstrate that, as already estimated in the main text of the paper, for the strains and dislocation densities considered, annihilation does not have a very significant influence on mobile dislocation density evolution over a single half cycle.

\end{document}